\documentclass[useAMS,usenatbib]{mnras}
\usepackage{times}
\usepackage{graphicx}
\usepackage{amsmath}
\usepackage{fmtcount}
\usepackage{amssymb}
\usepackage{gensymb}
\usepackage{natbib}
\usepackage{color}
\usepackage{booktabs}
\bibpunct{(}{)}{;}{a}{}{,}
\usepackage{psfig}
\title[The field/isolated lenticular galaxy NGC\,4546]{Field/Isolated lenticular galaxies with high $S_N$ values: the case of NGC\,4546 and its globular cluster system}

\author[Escudero et al.]{Carlos G. Escudero$^{1,2,3}$, Favio R. Faifer$^{1,2,3}$, Anal\'ia V. Smith Castelli$^{1,2,3}$, 
  \newauthor
  Mark A. Norris$^{4}$, Juan C. Forte$^{3,5}$ \\
  \\
$^{1}$Facultad de Cs. Astron\'omicas y Geof\'isicas, UNLP, Paseo del Bosque S/N, 1900 La Plata, Argentina \\
$^{2}$ Instituto de Astrof\'isica de La Plata (CCT La Plata - CONICET - UNLP)\\
$^{3}$ Consejo Nacional de Investigaciones Cient\'ificas y T\'ecnicas, Rivadavia 
1917, C1033AAJ Ciudad Aut\'onoma de Buenos Aires, Argentina \\
$^{4}$ Jeremiah Horrocks Institute, University of Central Lancashire, United Kingdom \\
$^{5}$ Academia Nacional de Ciencias de Buenos Aires, C1014AAE Ciudad Aut\'onoma de Buenos Aires, Argentina
}

\begin{document}

\pagerange{\pageref{firstpage}--\pageref{lastpage}} \pubyear{}
\maketitle

\label{firstpage}

\begin{abstract}

We present a photometric study of the field lenticular galaxy NGC\,4546 using Gemini/GMOS imaging in $g'r'i'z'$. We perform a 2D image decomposition of the surface brightness distribution of the galaxy using {\sc{galfit}}, finding that four components adequately describe it. The subtraction of this model from our images and the construction of a colour map allow us to examine in great detail the asymmetric dust structures around the galactic centre.
In addition, we perform a detailed analysis of the globular cluster (GC) system of NGC\,4546. Using a Gaussian Mixture Model algorithm in the colour-colour plane we detected hints of multiple groups of GC candidates: the classic blue and red subpopulations, a group with intermediate colours that present a concentrated spatial distribution towards the galaxy, and an additional group towards the red end of the colour distribution. 
We estimate a total GC population for NGC\,4546 of $390\pm60$ members and specific frequency $S_N=3.3\pm0.7$, which is relatively high compared to the typical value for galaxies of similar masses and environment. 
We suggest that the unusual GC population substructures were possibly formed during the interaction that led to the formation of the young ultra-compact dwarf (NGC~4546-UCD1) found in this system. Finally, we estimate the distance modulus of NGC\,4546 by analyzing its luminosity function, resulting in $(m-M)=30.75\pm0.12$ mag (14.1 Mpc).

\end{abstract}

\begin{keywords}
galaxies: individual -galaxies: elliptical and lenticular, cD -galaxies: star clusters
\end{keywords}

%=====================================================================
%============================================================================

\section{INTRODUCTION}
\label{sec:intro}

In the last decades, the study of globular cluster (GC) systems has become an important tool to obtain information about the formation history and evolution of early-type galaxies. This is because the origin of GCs seems to be closely linked to periods of intense star formation \citep[e.g.,][]{kissler98,kuntschner02,puzia04,puzia05,strader07,sesto18,forte19}. 

In general, GC systems associated with early-type galaxies located in different environments show bimodal colour distributions to a greater or lesser extent. These components are associated with two old \citep[$>$9 Gyr;][]{strader05} GC subpopulations which present different metallicities: the blue (or metal-poor) and the red (or metal-rich) clusters. The different features and properties shown by both groups \citep[see][]{brodie06} suggest different formation epochs and/or mechanisms \citep[e.g.,][]{bekki08,lee10,kruijssen14,kruijssen15,choksi18}. However, a handful of galaxies in the local Universe present complex GC systems showing multiple peaks in their colour distributions, possibly indicating star formation episodes at lower redshifts \citep{peng06,blom12,tonini13,caso13,caso15,escudero15,sesto16}. 
Therefore, the study of galaxies which display both classical and old GC subpopulations, alongside subpopulations with different metallicities and/or ages, provides important constraints for galaxy formation models as well as information about the star formation and assembly histories of individual galaxies. 

In this context, we focus on the photometric analysis of the GC system associated with the lenticular (S0) galaxy NGC\,4546, in order to characterize and inquire about the evolutionary history of the system as a whole. NGC\,4546 is a field \citep[$log\,(\rho)=-1.14$ Mpc$^{-3}$;][]{cappellari11} nearly edge-on galaxy, with only a single previous study of its GC system \citep{zaritsky15}. These authors, using images from the Spitzer Survey of Stellar Structures in Galaxies \citep[S$^4$G;][]{sheth10}, estimated the total population of the GC system within 50\,kpc as $N_\mathrm{tot}=120$ members. On the other hand, photometric and spectroscopic observations of the galaxy \citep{galletta87,bettoni91,emsellem04,kuntschner10,ricci15} revealed irregular dust lanes, pseudo spiral arms, and the presence of ionized gas rotating in the opposite direction to its stellar disk. 
In addition, \citet{norris11} and \citet{norris15} studied the stellar population of the ultra-compact dwarf (UCD) NGC\,4546-UCD1 linked to the NGC\,4546 system. These authors found that NGC\,4546-UCD1 displays an extended star formation history with a light-weighted age of $\sim4$ Gyr, suggesting that it is the remnant nucleus of a dwarf galaxy that was tidally disrupted by NGC\,4546. All these pieces of evidence point out that NGC\,4546 has experienced at least one minor merger event in the recent past.

Adopting the distance modulus $(m-M)_0=30.73\pm0.14$ for NGC\,4546 (see Table \ref{Table_1}) and the scale 0.067 kpc/arcsec, the effective radius of the galaxy ($R_\mathrm{eff}=22.23$ arcsec) is equivalent to $\approx1.5$ kpc projected distance. Aditional relevant parameters of the galaxy are shown in Table \ref{Table_1}.

The structure of the paper is as follows. In Section \ref{sec:data} we present our observational data, the reduction process and photometric calibration. In addition, this section gives the completeness tests performed on the data, as well as a description of the comparison field adopted in this work. Section \ref{sec:photo} describes the overall properties of the galaxy light, and the identification of the photometric sub-components of NGC\,4546; in Section \ref{sec:GC} we focus on the general analysis of the GC system, as well as the possible subpopulations that comprise it. Finally, a summary of the results and the conclusions of this work are presented in Section \ref{sec:summary}.

\begin{table}
\centering
\caption{Properties of NGC\,4546. Equatorial and galactic coordinates from NED (http://ned.ipac.caltech.edu/); morphological classification (NED); $V$ magnitude from \citet{ho11}; distance modulus from \citet{tully13}; distance in Mpc; effective radius and stellar velocity dispersion from \citet{cappellari13}; heliocentric velocity from \citet{cappellari11}; stellar mass from \citet{norris11}.}
\label{Table_1}
%\scriptsize
\begin{tabular}{lcc}
\toprule
\toprule
\multicolumn{1}{c}{\textbf{Property}} &
\multicolumn{1}{c}{\textbf{Value}} &
\multicolumn{1}{c}{\textbf{Units}} \\
\midrule
$\mathbf{\alpha}$            &  12:35:29.5      & h:m:s (J2000) \\
$\mathbf{\delta}$            &  $-$03:47:35.5   & d:m:s (J2000) \\
{\it l}                      &  295:13:38.5     & d:m:s   \\
{\it b}                      &  58:50:23.3      & d:m:s   \\
Type                         &  SB0$^{-}$(s)     & --      \\
V${}_{\mathbf T}^{\mathbf 0}$     &  10.57$\pm$0.01  & mag     \\
(m--M)$_0$                   &  30.73$\pm$0.14  & mag     \\
Dist.                        &  14.0$\pm$0.9    & Mpc     \\
$R_\mathrm{eff}$              &  22.23           & arcsec   \\
$\sigma_e$                   &  188             & km/s    \\
V$_\mathrm{hel}$.             &  1057$\pm$5      & km/s     \\
Stellar mass                 &  2.7$\times$10$^{10}$  & M$_\odot$ \\
\bottomrule
\end{tabular}
\end{table}

%============================================================================
%============================================================================

\section{DATA}
\label{sec:data}
\subsection{Observations and Data Reduction}
\label{sec:obs}
\begin{figure*}
\resizebox{0.95\hsize}{!}{\includegraphics{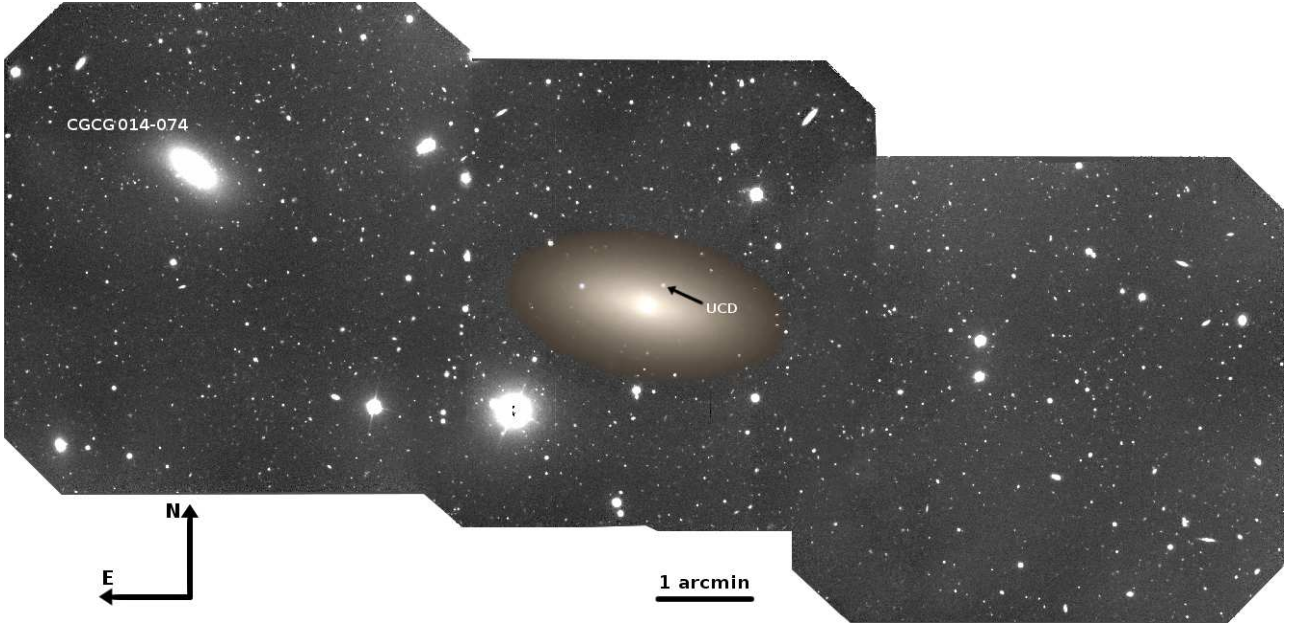}}
\caption{Mosaic created with the three GMOS fields used in this study. A false colour image of the galaxy was superimposed. In addition, the galaxy companion CGCG\,014-074 is observed towards the northeast. The black arrow shows the location of the UCD studied by \citet{norris11,norris15}. Considering the distance of 14$\pm$0.9 Mpc \citep{tully13} for NGC\,4546, 1\,arcsec=0.067 kpc.}
\label{figure_1}
\end{figure*}

The photometric study is based on a mosaic consisting of three deep fields (Figure \ref{figure_1}) obtained with the GMOS camera of the Gemini-South telescope \citep{hook04}. The images were taken under excellent seeing conditions ($0.46-0.71$ arcsec), as part of Gemini programmes GS-2011A-Q-13 (PI: Norris, M.) and GS-2014A-Q-30 (PI: Escudero, C.), using 4$\times$100 sec exposures in the $g'$,$r'$,$i'$ filters \citep{fukugita96} and 4$\times$290 sec exposures in the $z'$ filter. Figure \ref{figure_1} shows NGC\,4546 in the central region of the mosaic, and also the presence of the dS0 edge-on galaxy CGCG\,014-074 with similar redshift \citep[V$_\mathrm{hel}$=998$\pm$54 km/s;][]{colless03} to NGC\,4546, located at a projected distance of $\sim$5.5 arcmin ($\sim$22 kpc) towards the northeast. 

The reduction process of our observations was carried out by specific Gemini/GMOS routines within {\sc{iraf}}\footnote{IRAF is distributed by the National Optical Astronomical Observatories, which are operated by the Association of Universities for Research in Astronomy, Inc., under cooperative agreement with the National Science Foundation} (version V2.15), such as {\sc{gprepare}}, {\sc{gbias}}, {\sc{giflat}}, {\sc{gireduce}}, {\sc{gmosaic}}, {\sc{gifringe}} and {\sc{girmfringe}}. The raw data was corrected using appropriate bias and flat-field frames obtained from the Gemini Observatory Archive (GOA) as part of the standard GMOS baseline calibrations.
Within this reduction stage, a feature present in the $i'$ and $z'$ frames is the night-sky {\em fringing}, caused by thin-film interference effects in the CCD detectors. To subtract this effect from our science data, blank sky calibration images were used. These images, with exposure times of 300 sec, are obtained by the Gemini Observatory each semester. We downloaded five blank sky images from GOA, which were combined using the {\sc{gifringe}} task. Subsequently, this resulting frame was used to remove the fringing pattern in our data using the task {\sc{girmfringe}}.

As the final step, the reduced images for each field and filter were co-added using the IRAF task {\sc{imcoadd}}, obtaining final $g'$, $r'$, $i'$ and $z'$ images for the three fields.  

%============================================================================
%============================================================================

\subsection{Photometry}
\label{sec:phot}
In order to obtain a better detection, classification and subsequent photometry of GC candidates, it was necessary to model and subtract the sky background as well as the brightness due to the galaxy halo. 
To do this, we followed the same procedure described in \citet{faifer11} and \citet{escudero15,escudero18}, using a script that combines features of SE{\sc{xtractor}} software \citep{bertin96}, along with {\sc{iraf}} filtering tasks. In addition, the script provides a catalogue for each science field containing all the objects detected in the $g'r'i'z'$ filters. Since typical GCs have effective radii of several parsecs \citep{harris09a,puzia14} and given the adopted distance to NGC\,4546 (Table \ref{Table_1}), they are expected to be observed as unresolved sources in our GMOS images. Therefore, we used the stellarity index parameter ({\sc{class star}}) of SE{\sc{xtractor}} as a selection criterion to separate between resolved and unresolved objects using the boundary value of 0.5.

After performing the objects detection, we selected between 20-30 isolated, unresolved, and non-saturated sources uniformly distributed over the GMOS field of view (FoV), with the aim to model the point spread function (PSF) on each science field and filter. Using the {\sc{daophot}} package \citep{stetson87} within {\sc{iraf}}, we fit a Moffat\,25 model to the light distribution of the objects through the {\sc{psf}} task, since this provides lower errors than those obtained by using another analytical functions included in the task mentioned above (e.g., Gaussian and Moffat\,15 models). In addition, as usual for GMOS images, we found that allowing a quadratic spatial variation of the PSF model over the FoV gives the best results. 
Subsequently, the corresponding PSF empirical models were applied to each unresolved source detected in our images through the {\sc{allstar}} task. In addition, we performed aperture corrections to the PSF magnitudes using the {\sc{mkapfile}} task.

%============================================================================
%============================================================================

\subsection{Photometric Calibration}
\label{sec:ph_cal}
In order to transform the instrumental PSF magnitudes to the AB standard system, a standard star field (E5-b; $\alpha_\mathrm{J2000}$=12:05:25, $\delta_\mathrm{J2000}$=$-$45:38:11) was observed in the same night as our central science field (field 2 containing the galaxy). 
The instrumental magnitudes of the photometric standard stars in this field were obtained using the {\sc{phot}} task within {\sc{iraf}}. Since the standard field is not very crowded, we decided to use a large aperture of 3 arcsec of radius ($\sim$5\,FWHM), to avoid performing an aperture correction. 

The transformation of our instrumental magnitudes to the standard system was carried out using the following expression:
\begin{eqnarray}
\label{cero}
 m_{std} = m_{zero} + m_{inst} - K_{CP} (X-1)  + CT (m_1-m_2)_{std}
\end{eqnarray}
\noindent where $m_{std}$ are the standard magnitudes, m$_{zero}$ are the photometric zero-points, $m_{inst}$ are instrumental magnitudes, K$_\mathrm{CP}$ is the mean atmospheric extinction at Cerro Pach\'on, $X$ is the airmass of our observations, CT is the coefficient of the colour term and $(m_1-m_2)_{std}$ are the colours listed in the fourth column of Table \ref{Table_2}.
The mean magnitude zero-point for each filter of the standard field was determined following the procedure of \citet{jorgensen09} and \citet{escudero15}. On the other hand, owing to the low number of standard stars in our standard field, the values of the colour terms used here were those published in this last mentioned work. 

Once the magnitudes of the objects belonging to the central field of our mosaic were transformed using the expression \ref{cero}, the corresponding data to fields 1 and 3 were calibrated using common sources located in the overlapping regions between them and field 2. Finally, we applied the Galactic extinction coefficient values given by \citet{schlafly11} (fifth column in Table \ref{Table_2}).  

\begin{table}
\centering
\caption{Coefficients used in the transformation to the standard system. m$_\mathrm{zero}^{*}$: final values of the photometric zero-points used for the field 2. CT: values of colour terms obtained from \citet{escudero15}. A$_{\lambda}$: Galactic extinction coefficients for each filter determined by \citet{schlafly11}.}
%\scriptsize
\begin{tabular}{cllll}
\multicolumn{1}{c}{ } \\
\toprule
\toprule
\multicolumn{1}{c}{\textbf{Filter}} &
\multicolumn{1}{c}{\textbf{m$_\mathrm{zero}^{*}$}} &
\multicolumn{1}{c}{\textbf{CT}} &
\multicolumn{1}{c}{\textbf{(m$_1$-m$_2$)}} &
\multicolumn{1}{c}{\textbf{A$_{\lambda}$}} \\
\midrule
 $g'$ &  28.27$\pm$0.01 & ~0.08   & $(g'-r')$ & 0.112 \\
 $r'$ &  28.36$\pm$0.01 & ~0.03   & $(g'-r')$ & 0.077 \\
 $i'$ &  27.98$\pm$0.02 & -0.02   & $(r'-i')$ & 0.057 \\
 $z'$ &  26.84$\pm$0.01 & ~0.0    & $(i'-z')$ & 0.043 \\
\bottomrule
\end{tabular}
\label{Table_2}
\end{table}

%============================================================================
%============================================================================

\subsection{Completeness Test}
\label{sec:test}
In order to quantify the completeness of our photometric catalogue, we performed a series of completeness experiments on each science field. We added 200 artificial sources at intervals of 0.2 mag on the $g'$ band image, covering the range $21<g'_0<29$ mag (a total of 8000 sources). This procedure was carried out using the {\sc{starlist}} and {\sc{addstar}} tasks of {\sc{iraf}}. The first one was used initially to generate a sample of artificial objects randomly distributed over the fields. In order to simulate the radial distribution of GC candidates over the mosaic, a power law function was used in the science field that contains the galaxy, and uniform distribution in the remaining fields where the galaxy halo is weak. On the other hand, the second task adds the respective objects in the science images using the PSF model previously obtained (see Section \ref{sec:phot}). 
Subsequently, we applied the same script mentioned in Section \ref{sec:phot} to recover the artificial objects added in the images. 

\begin{figure}
\resizebox{0.95\hsize}{!}{\includegraphics{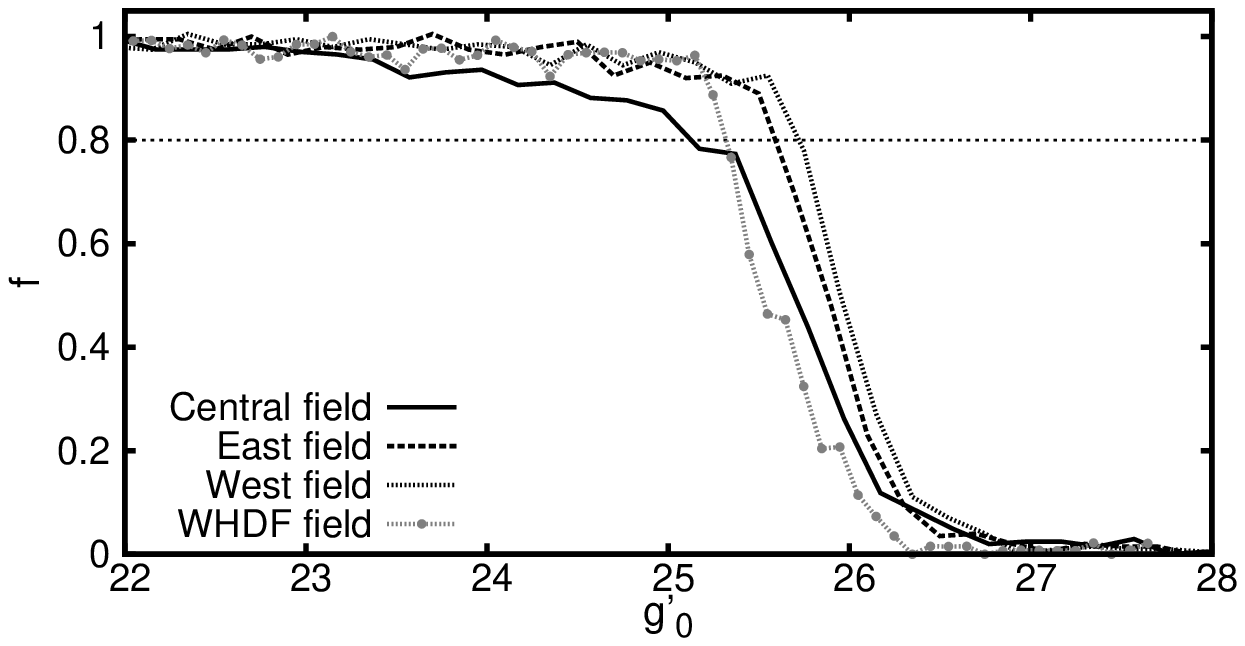}}
\centering
\resizebox{0.95\hsize}{!}{\includegraphics{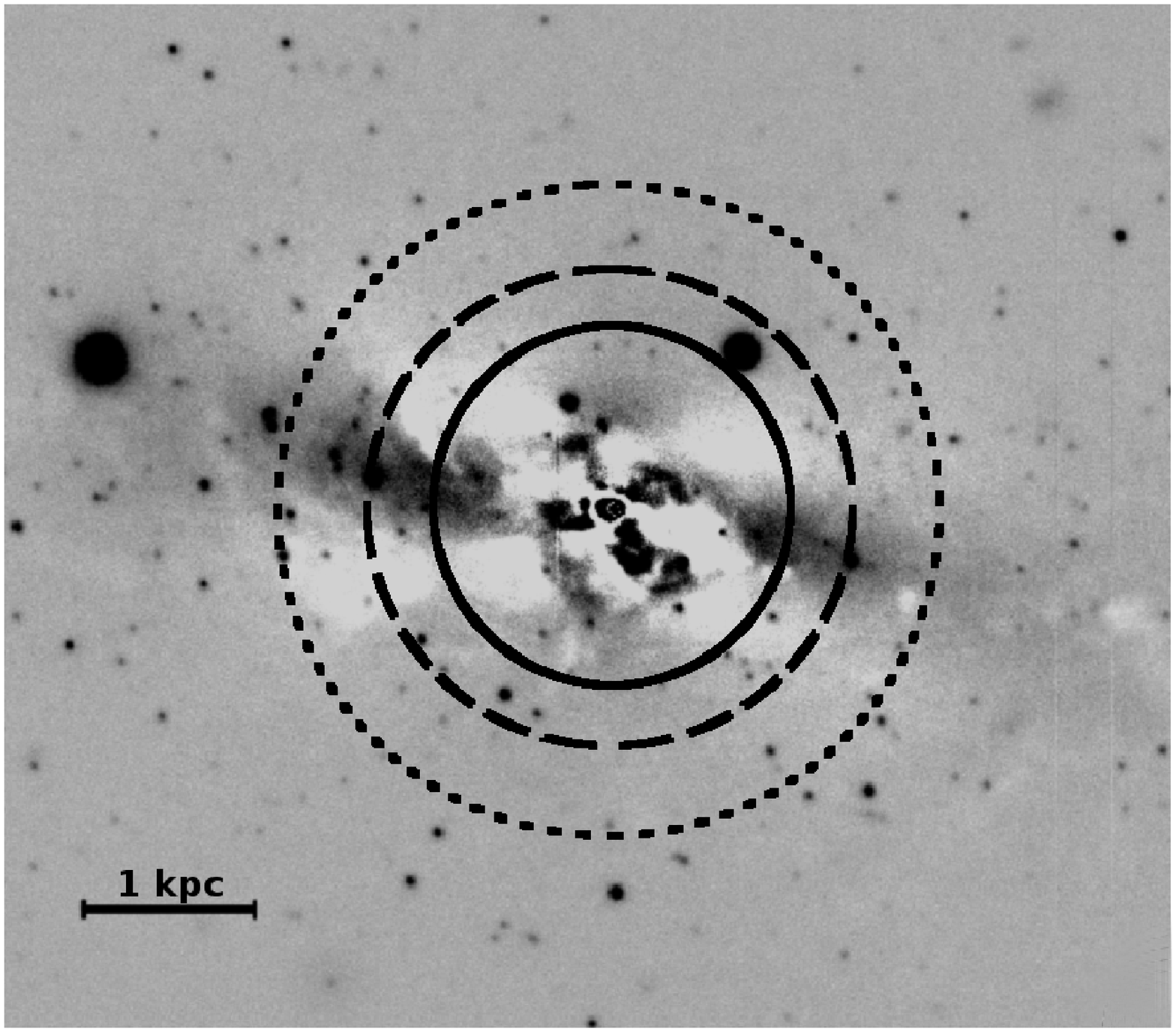}}
\caption{Upper panel: completeness fraction as a function of $g'_0$ magnitude. The solid, dashed and dotted lines correspond to the three studied GMOS fields of NGC\,4546. The dotted-dashed grey line shows the completeness of the comparison field (WHDF; Section \ref{sec:comp}). Bottom panel: image $g'$ of the central region of NGC\,4546 after the subtraction of its surface brightness. The black continuous, dashed and dotted circles indicate the regions from which a completeness level greater than 80 per cent is obtained for $g'_0=25$, 25.2 and 25.4 mag, respectively.}
\label{figure_2}
\end{figure}

The obtained results of our completeness tests are shown in Figure \ref{figure_2}. The upper panel shows the fraction of artificial objects recovered ($f$), as a function of the input magnitude. The same panel shows with the dotted-dashed line the completeness analysis performed over the comparison field used in this work (see Section \ref{sec:comp}).
As seen in Figure \ref{figure_2}, at $g'_0=25.2$ mag our data have a completeness level greater than 80 percent. 

In addition, we perform completeness tests as a function of the galactocentric radius, since lower completeness is expected towards the central region of the galaxy due to its brightness. The experiments were performed considering the magnitude value previously obtained ($g'_0=25.2$ mag), and values above and below it ($g'_0=25$ and $g'_0=25.4$ mag). The lower panel of Figure \ref{figure_2} shows the different limits in galactocentric radius (16, 22 and 30 arcsec) where an 80 per cent completeness level is obtained for the magnitudes $g'_0=25$, 25.2 and 25.4, respectively. 
These results were considered for the analysis of the azimuthal and radial distribution of the GC system of NGC\,4546 (Section \ref{sec:azimutal}), as well as for its luminosity function (Section \ref{sec:lumin}).

%============================================================================
%============================================================================

\subsection{Comparison Field}
\label{sec:comp}
Previous studies \citep[e.g.][]{bridges06,pierce06a,pierce06b,faifer11} have shown that observations in three or more photometric bands with good seeing conditions (FWHM$<$1 arcsec) allows selection of GC samples with limited contamination by background objects.
The main contamination sources in catalogues are associated with background galaxies, unresolved stellar systems, and Milky Way (MW) stars. Most of these objects can present colours within the typical ranges displayed by GCs \citep{fukugita95}. For this reason, it is of utmost importance to consider the effect that these objects can have on the integrated colour distributions, as well as on any other analysis performed on extragalactic GC systems.

\begin{figure}
\resizebox{0.95\hsize}{!}{\includegraphics{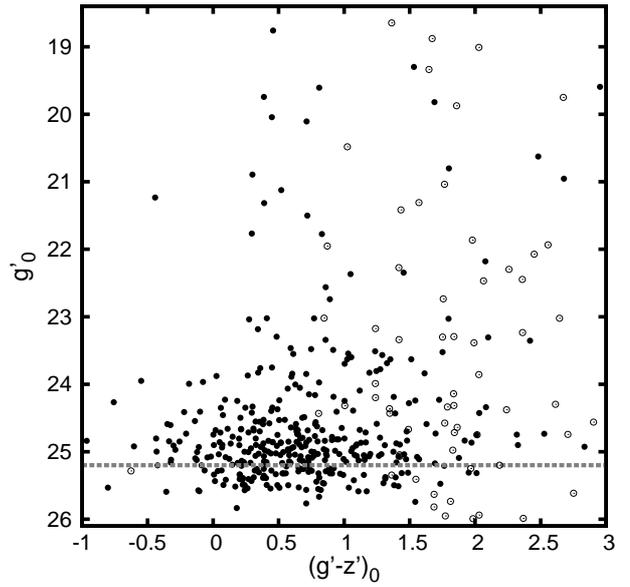}}
\caption{Colour-magnitude diagram with the unresolved sources detected in the WHDF field (black dots) and those obtained from {\sc{trilegal}} Galaxy model (open circles). Grey dashed line shows the limiting magnitude adopted in Section \ref{sec:test} ($g'_0=25.2$ mag) corresponding to 80 percent of completeness.}
\label{figure_3}
\end{figure}

We were unable to obtain a comparison field as part of our Gemini programme. Furthermore, a GOA search for imaging of similar depth and quality near to NGC\,4546 also gave a negative result. For these reasons, we decided to use the William Herschel Deep Field \citep[WHDF,][]{metcalfe01} to estimate the contamination in our photometric catalogue. The reduction process of this field was performed according to the procedures described in Sections \ref{sec:obs}-\ref{sec:ph_cal}. Used as comparison field in other works \citep{faifer11,caso15,salinas15}, this dataset has the advantage that was observed with the GMOS instrument in the $g'r'i'z'$ filters with a photometric depth and FWHM comparable to our data (upper panel of Figure \ref{figure_2}). In addition, according to the Galactic latitude and longitude of this field ($b=-61.7^\circ$, $l=107.5^\circ$), the largest number of contaminants would be due to unresolved background galaxies \citep{faifer11}. Therefore, we complemented this sample with the expected stars in the Galactic region of NGC\,4546, using the stellar population synthesis code {\sc{trilegal}} \citep{girardi05}. Based on this analysis, we estimate a contamination of 11\% in our GC sample for $g'_0<25.2$ mag.
Figure \ref{figure_3} shows the colour-magnitude diagram with the objects detected in the WHDF field identified as unresolved sources according to SE{\sc{xtractor}} ({\sc{class star}}$>$0.5), and those obtained by {\sc{trilegal}}.

%============================================================================
%============================================================================

\section{Surface Photometry of NGC\,4546}
\label{sec:photo}
We analyzed the light distribution of NGC\,4546 in the $g'$ and $z'$ images, in order to obtain the surface brightness profiles of the galaxy and also information about its different photometric structures. To do this, we ran the IRAF task {\sc{ellipse}} commonly used to measure the surface photometry of galaxies \citep{yuli11,lane13,escudero15,escudero18}, and the {\sc{galfit}}\footnote{https://users.obs.carnegiescience.edu/peng/work/galfit/galfit.html} 3.0.5 software \citep{peng02,peng10} to perform a 2D multi-component decomposition. Furthermore, the results obtained from these analyses were subsequently used to compare with some features exhibited by the GC system of NGC\,4546 (see Section \ref{sec:azimutal}).

Given that the galaxy's light fills the GMOS FoV of the central pointing, in order to estimate a fair value of the sky background level and perform a good modelling of the galaxy's profile at large radii, we build $g'$ and $z'$ mosaics with our images using the THELI\footnote{https://www.astro.uni-bonn.de/theli/} data reduction pipeline \citep{erben05,schirmer13}. This allows us to reach a distance of 7.4 arcmin ($\sim$30 kpc) from the galactic centre to the edge of the mosaic.

\subsection{Ellipse Model}
\label{sec:ellipse}
Initially, we ran {\sc{ellipse}} in the $g'$-band to obtain the surface brightness profile of the galaxy, previously masking unresolved and extended bright sources in the mosaic using the {\sc{iraf}} task {\sc{objmask}}. In addition, since the galactic centre in our images is saturated ($r<1.5$ arcsec), we also mask this region. During the fit process, we allow the free variation of the isophotal parameters (centre, ellipticity, and position angle ($PA$) of the ellipses) in the range of 0.16 to 2.5 arcmin ($0.64-10.05$ kpc) of semi-major axis (SMA). From this last value and up to SMA$\sim7$ arcmin (28.1 kpc), given the low surface brightness of the galaxy in these regions, all the parameters mentioned above were fixed in order to ensure the convergence of the fitting process. The same procedure was subsequently applied to the $z'$ mosaic.

\begin{figure}
\resizebox{0.95\hsize}{!}{\includegraphics{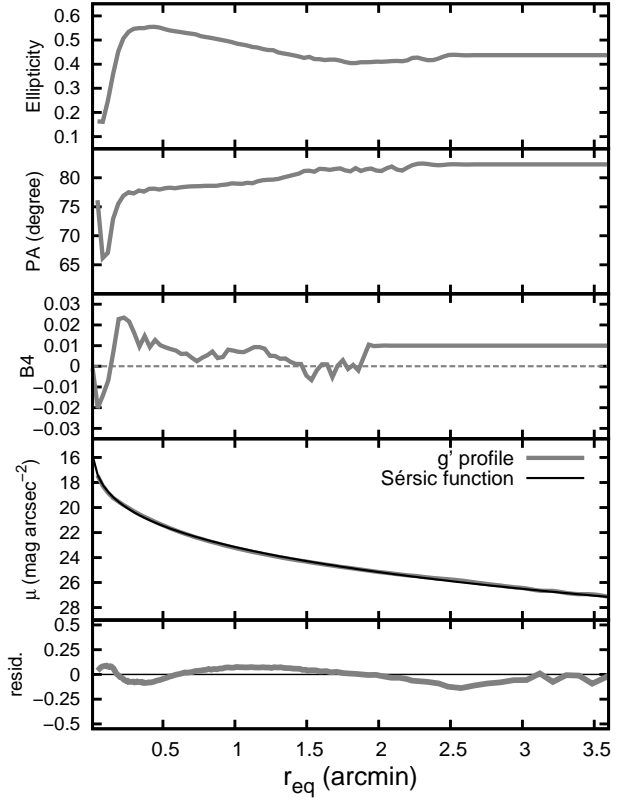}}
\caption{Isophotal parameters as a function of the equivalent galactocentric radius. From top to bottom, the different panels show the ellipticity, position angle, Fourier coefficient $B_4$, surface brightness profile in $g'$ filter, and the residual between the data and the fitted S\'ersic function.}
\label{figure_4}
\end{figure}

Figure \ref{figure_4} shows the variation of the isophotal parameters (ellipticity, $PA$ and the Fourier coefficient $B_4$) and the surface brightness profile ($g'$ filter) as a function of the equivalent radius ($r_\mathrm{eq}$\footnote{$r_\mathrm{eq}=a\sqrt{1-e}$, with $a$ and $e$ the semi-major axis and ellipticity of the ellipses}). As can be seen, the isophotal fit in the central region of NGC\,4546 shows a significant variation in the $PA$, changing $\sim15^\circ$ within $r_\mathrm{eq}<0.2$ arcmin (0.8 kpc). This variation is also reflected in the ellipticity, ranging from $\sim0.15$ to 0.5 within the mentioned radius.
In addition, the Fourier coefficient $B_4$ changes from boxy ($B_4<0$) to discy ($B_4>0$) isophotes in this same region. 

\begin{figure}
\centering
\resizebox{0.96\hsize}{!}{\includegraphics{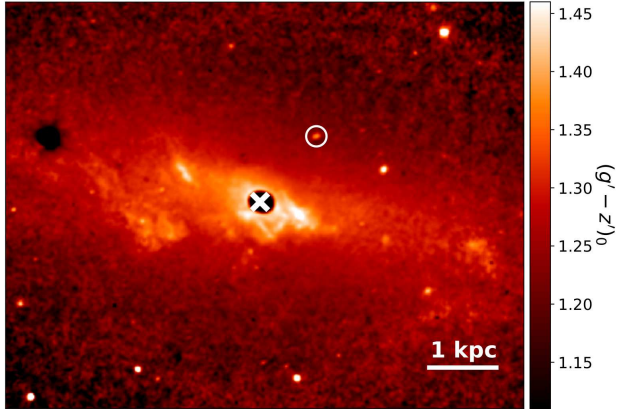}}
\caption{$(g'-z')_0$ colour map of NGC\,4546. Irregular dust structures can be observed around the galactic centre. The white cross and circle indicate the location of the galactic centre and the UCD, respectively.}
\label{figure_5}
\end{figure}

After subtracting our $g'$ model, different dust regions around the galaxy's nucleus are revealed. Some of these structures extend up to a distance of 0.86 arcmin (3.5 kpc) from the galactic centre. 
To highlight these structures and any other peculiarities in the galaxy, we generated a 2D colour map using our $g'$ and $z'$ images. 
The colour map (Figure \ref{figure_5}) reveals in great detail the existence of remarkable irregular dust structures along the major axis of NGC\,4546, with colours ranging from $(g'-z')_0\sim1.3$ to 1.5 mag, and also a smooth colour gradient of the stellar component of the galaxy, of $\Delta(g'-z')_0/\Delta r \sim0.1$ mag within $r_\mathrm{eq}<0.8$ arcmin ($\sim$3 kpc).
In addition, the location of the UCD \citep[NGC\,4546-UCD1;][]{norris15} is shown with a white circle in this figure.

In order to characterize the global surface brightness of the galaxy and obtain their structural parameters, we fit the S\'ersic function \citep{sersic68} to the obtained profiles $g'$ and $z'$. The analytical expression of this function has the following form:
\begin{equation}\label{eq:sersic}
\mu(r)= \mu_\mathrm{eff} + \left(\frac{2.5\,b_n}{\ln 10}\right)\left[\left(\frac{r}{R_\mathrm{eff}}\right)^{(1/n)}-1\right],
\end{equation}
with $R_\mathrm{eff}$ the effective radius of the galaxy, $\mu_\mathrm{eff}$ the surface brightness at that radius, $n$ the S\'ersic index, and $b_n=1.9992\,n-0.3271$. 
The fits were performed in the range of 0.04 to 4.5 arcmin, with the aim to avoid the saturated innermost region, and the outer region probably influenced by the sky level. The parameters obtained are listed in Table \ref{Table_3}.
The mean value of the effective radius obtained for NGC\,4546 of the fits is $\langle R_\mathrm{eff} \rangle=21.1\pm0.8$ arcsec ($\sim1.39\pm0.05$ kpc). This value is in good agreement in comparison with the values obtained by \citet{bettoni91} ($R_\mathrm{eff}=21.7$ arcsec) and \citet{cappellari13} ($R_\mathrm{eff}=22.23$ arcsec).
In addition, we compute the total $g_0'$ and $z_0'$ magnitudes of the galaxy by integrating expression \ref{eq:sersic}. 

\begin{table}
\centering
\caption{Best-fit parameters of the S\'ersic profile ($\mu_\mathrm{eff}$, $R_\mathrm{eff}$ and $n$) and total magnitudes obtained in filters $g'$ and $z'$.
} 
\label{Table_3}
\begin{tabular}{cllll}
\toprule
\toprule
\multicolumn{1}{c}{\textbf{Filter}} &
\multicolumn{1}{c}{\textbf{$\mu_\mathrm{eff}$}} &
\multicolumn{1}{c}{\textbf{$R_\mathrm{eff}$}} &
\multicolumn{1}{c}{\textbf{$n$}} &
\multicolumn{1}{c}{\textbf{$m_0$}}\\
\multicolumn{1}{c}{\textbf{}} &
\multicolumn{1}{c}{\text{(mag\,arcsec$^{-2}$)}} &
\multicolumn{1}{c}{\text{(arcsec)}} &
\multicolumn{1}{c}{\textbf{}} &
\multicolumn{1}{c}{\text{(mag)}} \\
\midrule
$g'$ &  20.84$\pm$0.05  & 22.20$\pm$0.6  & 4.06$\pm$0.08 & 10.71$\pm$0.11 \\
$z'$ &  19.26$\pm$0.04  & 19.93$\pm$0.6  & 3.87$\pm$0.07 & 9.40$\pm$0.10 \\
\bottomrule
\end{tabular}
\end{table}

%%%%%%%%%%%%%%%%%%%%%%%%%%%%%%%%%%%%%%

\subsection{Galfit Model}
\label{sec:galfit}
Although the 1D surface brightness profile of NGC\,4546 is modelled relatively well with only one S\'ersic function (Section \ref{sec:ellipse}), this approach is not very effective in identifying the different photometric subcomponents of the galaxy. Therefore, we used the software {\sc{galfit}} to perform 2D multi-component decomposition in our images. {\sc{galfit}} is a 2D data analysis algorithm designed to fit multiple parametric functions simultaneously (S\'ersic, Gaussian, Moffat, etc.), and also allow the modelling of non-axisymmetrical structures (bars, spiral arms, tidal features, etc.).

In this work, we ran {\sc{galfit}} on our $g'$ and $z'$ mosaics and fit S\'ersic and exponential functions, given their flexibility. In addition, since the galaxy presents boxy and discy isophotes (see Section \ref{sec:ellipse}), in order to quantify the degree of asymmetry in the galaxy and simulate these types of isophotes, we considered the Fourier modes $m=1$ and $m=4$. 
Initially, we started by fitting a single S\'ersic component using as initial parameters those obtained in Section \ref{sec:ellipse}. Afterwards, we added additional components and carefully inspecting the residuals obtained in each case. In each run, we considered the object mask, and the empirical PSF previously obtained in Section \ref{sec:phot}. Since the galaxy centre in our images is saturated ($r<1.5$ arcsec), we also mask this region before running {\sc{galfit}}. Several tests were performed by varying the initial values of the parameters of each component, in order to test the reliability of our results. 
In all cases, these modifications did not have significant effect on the final models.

The maximum number of fitted components was determined following the procedure of \citet{huang13}. The 1D surface brightness profile of each obtained model was extracted, and subsequently, the isophotal parameters were compared with that derived from the original data. According to these authors, ``{\it the best model is one that it contains a minimum number of components with reasonable, robust parameters that describe visibly distinct structure''}. 

\begin{figure}
\centering
\resizebox{0.95\hsize}{!}{\includegraphics{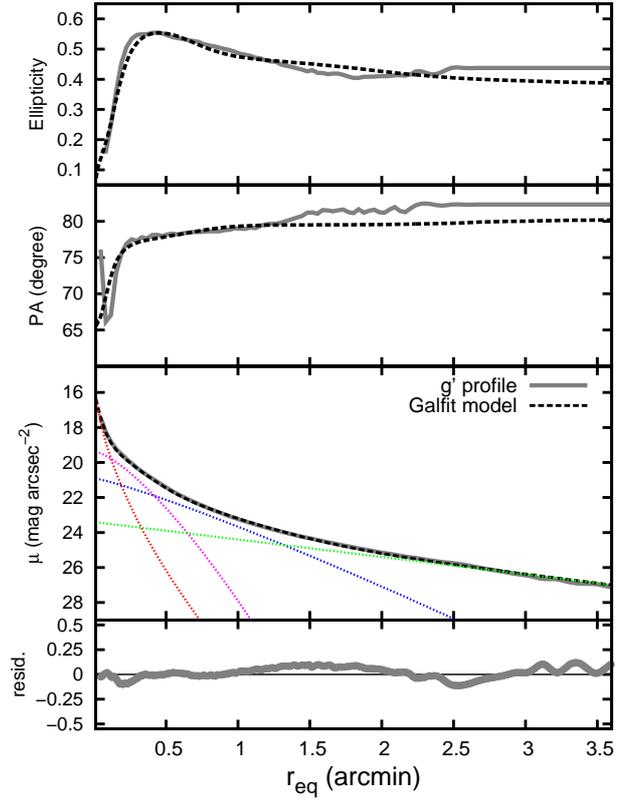}}
\resizebox{0.95\hsize}{!}{\includegraphics{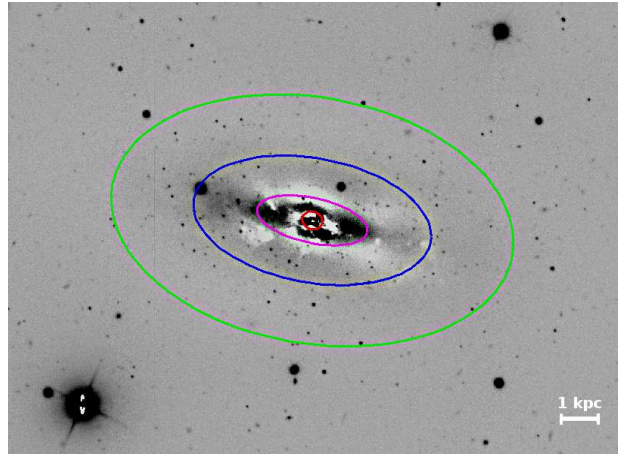}}
\caption{Upper panel: comparison between the 1D profile of the data and the obtained {\sc{galfit}} model. From top to bottom, the different panels show the ellipticity, position angle, surface brightness profile in $g'$ filter, and the residual between the data and the four considered functions. The red, magenta, blue and green dotted lines show the different subcomponents used for the {\sc{galfit}} model. Bottom panel: residual $g'$-image after subtracting the obtained {\sc{galfit}} model. The ellipses represent $R_\mathrm{eff/s}$, ellipticity, and $PA$ of each fitted component (colour figure in the online version).}
\label{figure_6}
\end{figure}

In our case, the best model (Figure \ref{figure_6}) in the aforementioned filters was obtained considering four components: three S\'ersic functions and one exponential. 
The innermost S\'ersic indicates the presence of the galaxy's bulge, while the remaining two S\'ersic components would indicate the presence of lens/disks. The fit of these three components is similar to the analysis obtained by \citet{gao18} on NGC\,4546. In that work, the authors mention that the galaxy is probably composed of a thin and a thick disk, denoted as a possible lens and a disk. In our analysis, the fit of these three functions left an excess of light in the residual image. Therefore, we consider a fourth additional model component through an exponential function, which would represent a faint extended halo. Table \ref{Table_4} lists the best-fit parameters for the final model in the filters $g'$ and $z'$. 
Although the bulge + disk decomposition performed by \citet{bettoni91} suggested the possible presence of a small bar in NGC\,4546 together with two spiral-like structure, we have not detected these features in our analysis. Therefore, our results are in agreement with \citet{gao18}.
\begin{table*}
\centering
%\scriptsize
\begin{tabular}{lccccccccc}
\multicolumn{10}{c}{}\\
\toprule
\toprule
\multicolumn{1}{c}{\textbf{Filter}} &
\multicolumn{1}{c}{\textbf{Model}} &
\multicolumn{1}{c}{\textbf{Component}} &
\multicolumn{1}{c}{\textbf{$\mu_\mathrm{eff/s}$}} &
\multicolumn{1}{c}{\textbf{$R_\mathrm{eff/s}$}} &
\multicolumn{1}{c}{\textbf{$n$}} &
\multicolumn{1}{c}{\textbf{$Ellip$}} &
\multicolumn{1}{c}{\textbf{$PA$}} &
\multicolumn{1}{c}{\textbf{$m_{g'/z'}$}} &
\multicolumn{1}{c}{\textbf{$\chi^2/\nu$}} \\
\multicolumn{3}{c}{} &
\multicolumn{1}{c}{(mag\,arcsec$^{-2}$)} &
\multicolumn{1}{c}{(arcsec)} &
\multicolumn{1}{c}{} &
\multicolumn{1}{c}{} &
\multicolumn{1}{c}{(degrees)} &
\multicolumn{1}{c}{(mag)} &
\multicolumn{1}{c}{} \\
\midrule
$g'$   &  S\'ersic  & Bulge     & 18.4  & 4.3   & 1.70  & 0.16  & 65.3  & 12.28   & 0.9   \\
       &  S\'ersic  & Lens/Disk & 20.5  & 23.1  & 0.72  & 0.60  & 77.5  & 11.13   &       \\
       &  S\'ersic  & Disk      & 22.4  & 49.5  & 0.86  & 0.48  & 79.6  & 11.30   &       \\
       &  Exp.      & Halo      & 23.4  & 83.6  & ----  & 0.39  & 80.1  & 11.10   &       \\
\hline
$z'$   &  S\'ersic  & Bulge     & 17.0  & 4.0   & 1.75  & 0.14  & 55.3  & 11.42   & 2.0   \\
       &  S\'ersic  & Lens/Disk & 18.9  & 22.1  & 0.94  & 0.59  & 76.9  & 9.51    &       \\
       &  S\'ersic  & Disk      & 21.3  & 45.9  & 0.84  & 0.46  & 80.4  & 10.38   &       \\
       &  Exp.      & Halo      & 21.9  & 61.0  & ----  & 0.36  & 76.9  & 10.27   &       \\
\bottomrule
\end{tabular}
\caption{Parameters obtained by {\sc{galfit}} for each fitted function. The different columns indicate the used filters, fitted models, the individual components, surface brightness for S\'ersic and exponential functions, effective radius ($R_\mathrm{eff}$) for the S\'ersic function and scale length ($R_\mathrm{s}$) for the exponential function, S\'ersic index, ellipticity, position angle, the apparent magnitude of each component in the respective filter, and the $\chi^2/\nu$ value obtained, respectively.}
\label{Table_4}
\end{table*}

As observed in Figure \ref{figure_6} (upper panel), the final model reproduces in great detail the global structure (ellipticity and $PA$) of the galaxy. The bottom panel in this figure shows the resulting image in filter $g'$ after subtracting the {\sc{galfit}} model, where the different regions of dust can be observed as whiter regions. 

%============================================================================
%============================================================================

\section{Globular Cluster System of NGC\,4546}
\label{sec:GC}
\subsection{Globular Cluster Candidates}
\label{sec:color}
\begin{figure}
\centering
\resizebox{0.99\hsize}{!}{\includegraphics{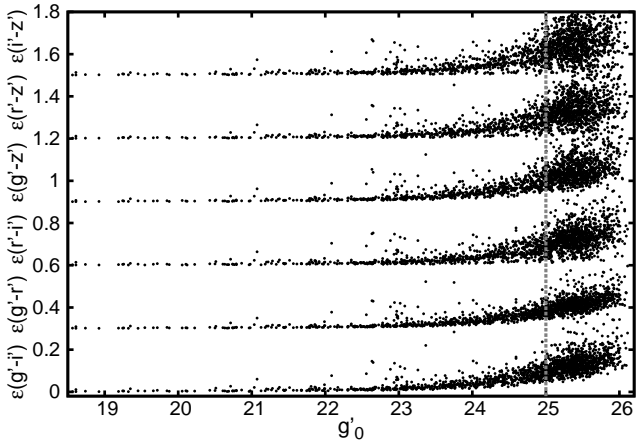}}
\caption{Photometric errors for different colour indices used in this study (shifted vertically to avoid overlapping). The vertical dashed line indicates the magnitude corresponding to the photometry errors for the colour indices of $\sim0.1$ mag.}
\label{figure_7}
\end{figure}

The quality of our photometry is reflected in Figure \ref{figure_7}, where the colour errors obtained at $g'_0=25.2$ mag (80\% completeness) are $\sim0.12$ mag. In order to reduce the contamination effect and the photometric errors in our catalogue, we adopted a slightly brighter magnitude limit of $g'_0=25$ mag. At this limit, the colour errors are lower than $\sim0.1$ mag.

\begin{figure}
\centering
\resizebox{0.88\hsize}{!}{\includegraphics{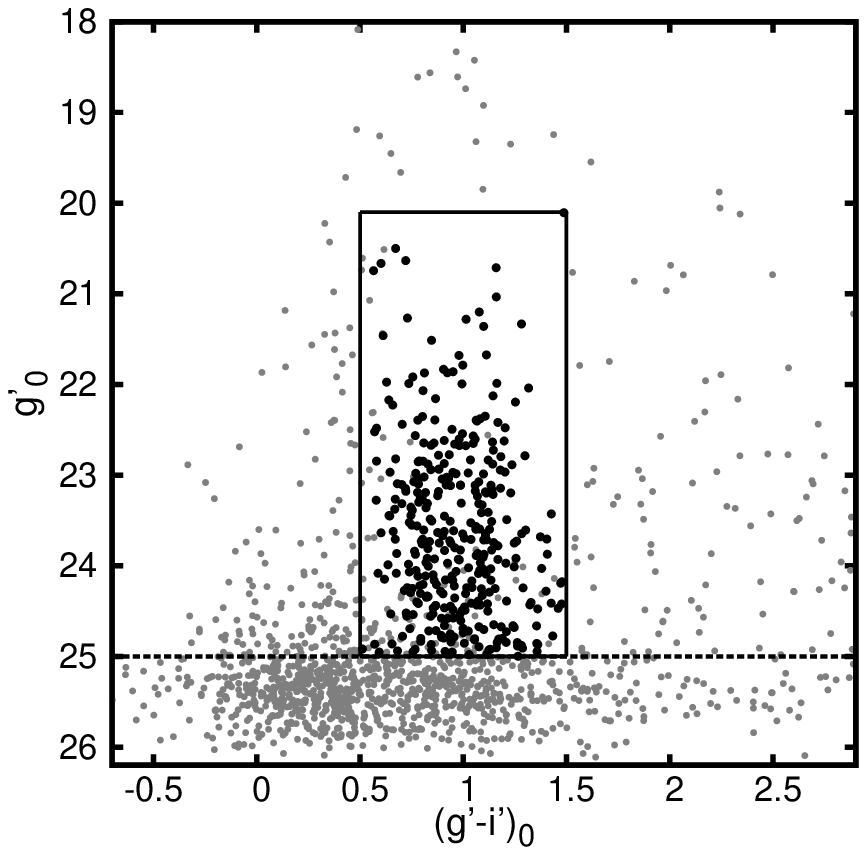}}
\resizebox{0.9\hsize}{!}{\includegraphics{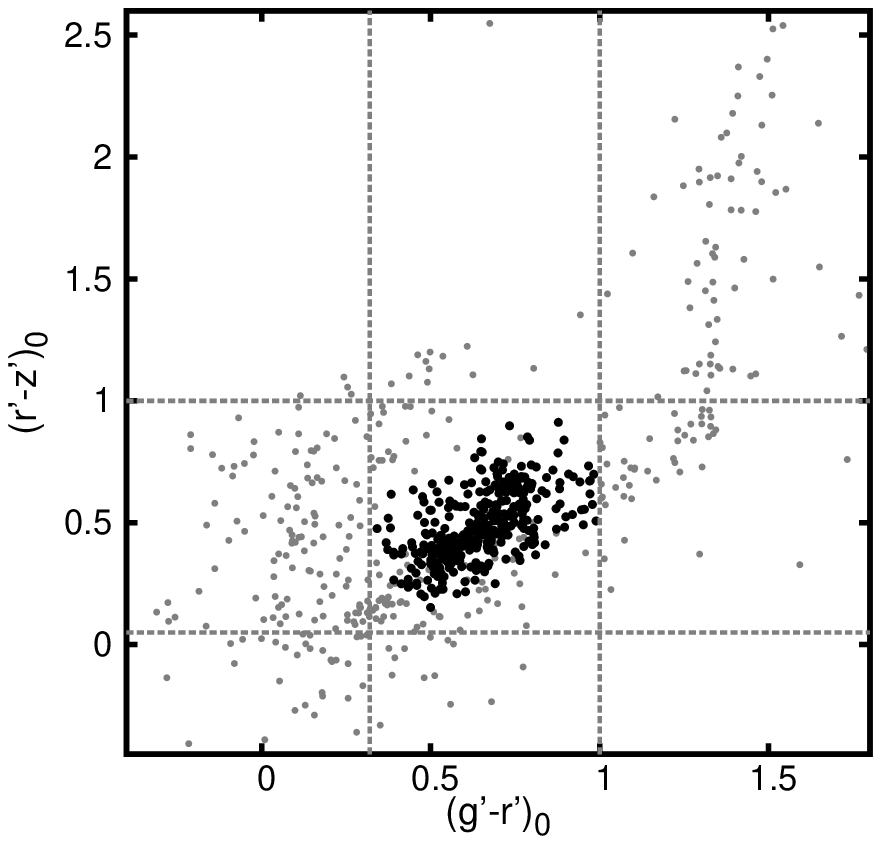}}
\caption{Upper panel: colour-magnitude diagram with all the unresolved sources detected in our GMOS mosaic (grey dots), as well as our final sample of GC candidates (black dots inside the box). The horizontal dashed line indicates the magnitude corresponding to $\epsilon(g'-i')\sim0.1$ mag. Bottom panel: colour-colour diagram with the unresolved sources brighter than $g'_0=25$ mag. The dashed grey lines show the colour ranges considered for the selection of GC candidates.}
\label{figure_8}
\end{figure}

Figure \ref{figure_8} shows the colour-magnitude (CM; upper panel) diagram and colour-colour (CC; bottom panel) diagram of all unresolved sources detected in our GMOS mosaic. The horizontal dashed line in the CM diagram indicates the limiting magnitude corresponding to $\epsilon_{(g'-i')}\sim0.1$ mag. As seen in the CC diagram, the GC candidates of NGC\,4546 are relatively easy to identify since they are grouped around specific colours ($(g'-r')_0\sim0.65$ mag, $(r'-z')_0\sim0.5$ mag; bottom panel in Figure \ref{figure_8}).

The selection of GC candidates was performed considering broad colour ranges similar to the values commonly used in this type of study and photometric system \citep{faifer11,escudero18}. The complex photometric substructure evidenced by our analysis presented in Section \ref{sec:galfit}, added to the irregular dust structures, the ionized gas rotating in the opposite direction to its stellar disk, and the young UCD, point to a merger event in the past of NGC 4546. In that context, and in order to detect not just ``classic'' blue and red GCs candidates, we decided to apply slightly wider colour cuts in our GC candidates selection. Therefore, we considered the following ranges $0.32<(g'-r')_0<1.0$, $0<(r'-i')_0<0.7$, $0.5<(g'-i')_0<1.5$, $0.6<(g'-z')_0<1.9$, $0.05<(r'-z')_0<1.0$, $-0.2<(i'-z')_0<0.7$ mag. Regarding the bright end in the CM diagram, we adopt the magnitude limit $g'_0=20.1$ mag, equivalent to the value $M_V\sim-11$ mag suggested by \citet{mieske06b} to separate GCs from massive clusters, UCDs, and/or MW stars.

As a final step, we conducted a visual inspection of the selected objects to avoid incorrect detections or artefacts in the final sample. A total of 350 GC candidates meet the criteria mentioned above, which are shown with black dots in Figure \ref{figure_8}.

%============================================================================
%============================================================================

\subsection{GC Colour Distribution}
\label{sec:histo}
\begin{figure}
\resizebox{0.88\hsize}{!}{\includegraphics{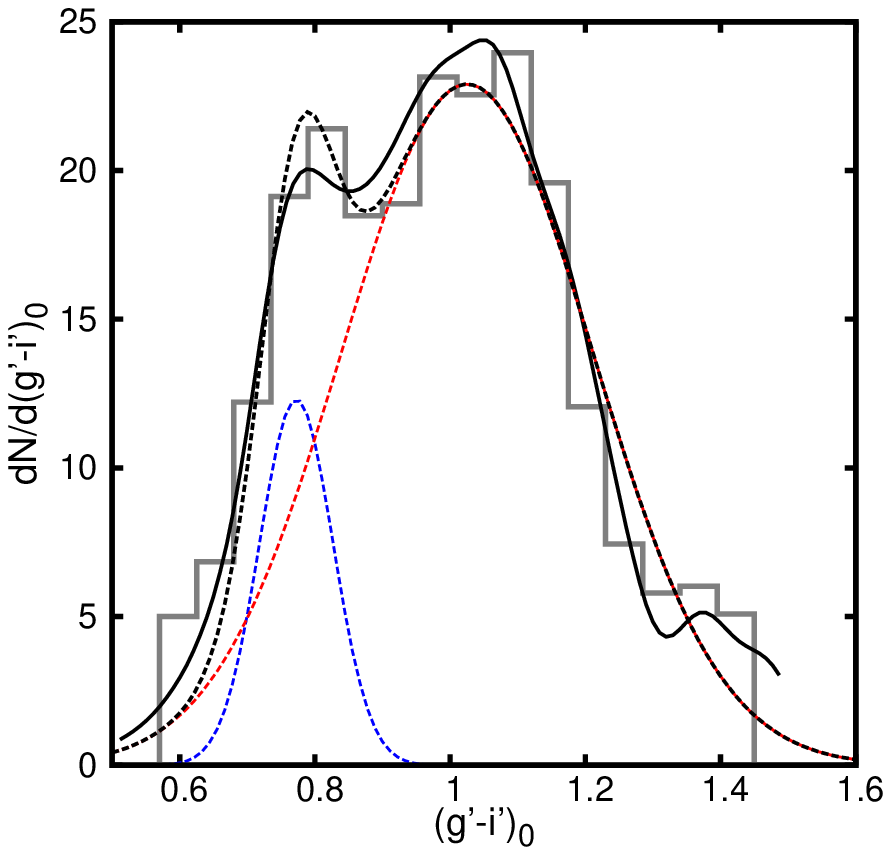}}
\resizebox{0.88\hsize}{!}{\includegraphics{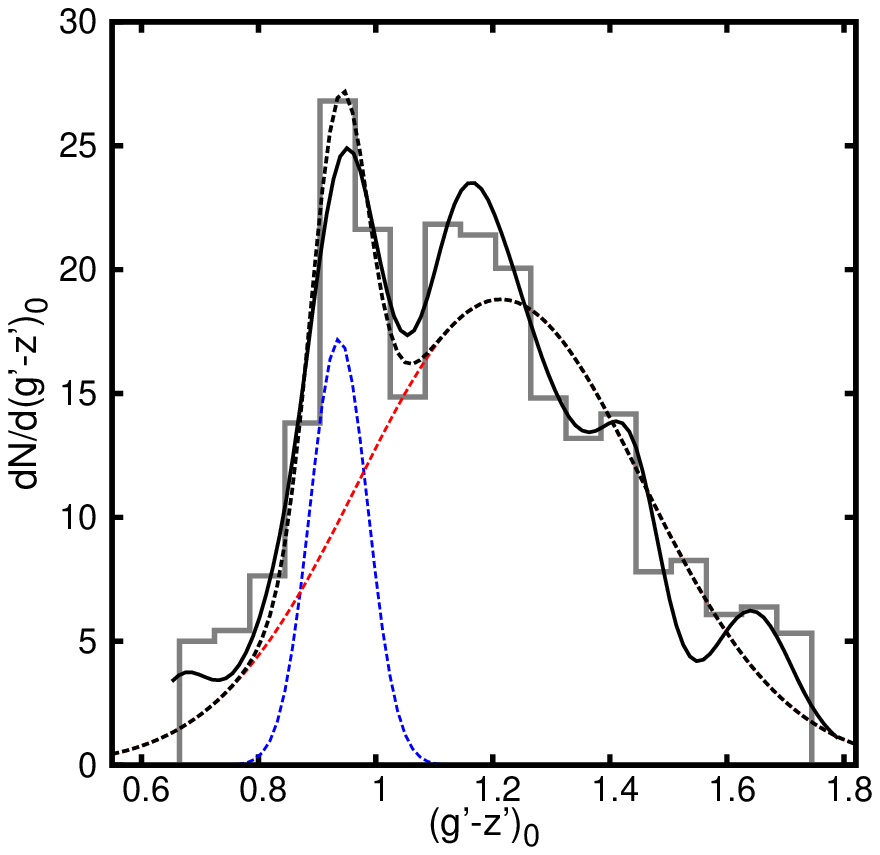}}
\caption{Upper panel: $(g'-i')_0$ background-corrected colour histogram for the GC candidates (grey line). The black line shows the smoothed colour distribution with a Gaussian kernel of 0.04 mag. Bottom panel: $(g'-z')_0$ background-corrected colour histogram (grey line). A Gaussian kernel of 0.06 mag was applied (black line). Blue and red dashed lines indicate the Gaussian components fitted by {\sc{rmix}}, where their sum is indicated by a black dashed line (colour figure in the online version).}
\label{figure_9}
\end{figure}

Initially, with the aim to identify the possible different GC subpopulations present in NGC\,4546, we obtained the background-corrected colour histogram $(g'-i')_0$ adopting a bin size of 0.055 mag (upper panel in Figure \ref{figure_9}). 
In addition, a Gaussian kernel with $\sigma=0.04$ mag was applied to obtain the smoothed colour distribution.
As seen in the figure, the histogram shows hints of different substructures that survive the background correction: at least two peaks at $(g'-i')_0\sim0.79$ and $(g'-i')_0\sim1.05$ mag, and a small group of GC candidates towards redder colours at $(g'-i')_0\sim1.4$ mag. The location of the first two peaks agrees with typical values for the blue and red GC subpopulations, respectively. 

Subsequently, we used the more metallicity-sensitive colour index $(g'-z')_0$ to corroborate the presence of these different groups. In this case, we built up the background-corrected $(g'-z')_0$ histogram adopting a slightly larger bin size of 0.06 mag and a Gaussian kernel with $\sigma=0.05$ mag (bottom panel in Figure \ref{figure_9}). As can be seen in the figure, two clear peaks are observed at $(g'-z')_0\sim0.9$ and $(g'-z')_0\sim1.15$ mag, with an extended count of objects towards redder colours. Using the relationship of \citet{peng06} about the behaviour of the blue and red GC subpopulations as a function of galaxy luminosity, we observe that the peak at $(g'-z')_0\sim0.9$ mag in NGC\,4546 results in good agreement with the values of blue subpopulations, while the peak at $(g'-z')_0\sim1.15$ mag is bluer compared with the typical values for red subpopulations.

In order to corroborate the bimodal or multimodal character of both colour distribution, as usual, we performed fits of different Gaussian components to the $(g'-i')_0$ and $(g'-z')_0$ background-corrected histograms using the RMIX\footnote{{\sc{rmix}} is publicly available at http://www.math.mcmaster.ca/peter/mix/mix.html} software. It is important to mention that although there is no physical reason to use Gaussian function in this type of analysis, they provide an adequate description of the integrated colour distributions and it allow us to compare NGC\,4546 with other galaxies studied in the literature. 

  The solutions obtained considering the fit with two Gaussian components (Figure \ref{figure_9}) were: $\mu_1=0.77\pm0.03$ ($\sigma_1=0.05\pm0.03$) and $\mu_2=1.02\pm0.03$ ($\sigma_2=0.18\pm0.02$) for the colour index $(g'-i')_0$; and $\mu_1=0.93\pm0.02$ ($\sigma_1=0.05\pm0.02$) and $\mu_2=1.21\pm0.03$ ($\sigma_2=0.24\pm0.02$) for the colour index $(g'-z')_0$. This analysis indicates that according to statistical indicators such as $\chi^2$, the bimodal description is appropriate in both cases. Our attempts of including a greater number of Gaussian components show that, although they provide a better formal description of the substructure of the distributions, they overfit the data and turn out not to be statistically significant. However, a visual inspection of both figures shows that if the $(g'-i')_0$ and $(g'-z')_0$ distributions are truly bimodal, then the red subpopulation is strangely and strongly dominant in NGC\,4546 (see the discussion in Section \ref{sec:summary}). Similar cases can be seen in \citet{blom12} and \citet{escudero15} where the presence of more than two subpopulations is proposed. 

Therefore, we explore a different approach consisting in evaluate the hypothesis of multimodality of the colour distribution in the colour plane. If those distributions in a plane that involves two metallicity-sensitive indices are constituted by two dominant subpopulations (the classic blue and red), in addition to other less populated ones, then its projection on one of the axes of the plane can make it more difficult to identify them. Especially, if one considers that NGC\,4546 does not have thousands of GCs as more massive galaxies.

In this context, using the joint information of the colour indices $(g'-z')_0$ and $(g'-i')_0$ via the CC diagram, we made a probabilistic classification of the different GC groups in NGC\,4546. To do this, we applied the Gaussian mixture model (GMM) algorithm using the open source machine learning library {\sc{scikit-learn}}\footnote{http://scikit-learn.org} for python \citep{desouza17}. GMM is a parametric probability density function to describe the data distribution in certain feature space as a weighted sum of Gaussian component densities. 

For a total of $K$ components in an $n$-dimensional parameter space, the final probability distribution $p(x)$ is given by the equation,
\begin{equation}\label{eq:gmm_px}
p(x)  = \sum_{i=1}^K w_i\phi(x;\vec{\mu}_i,\vec{\Sigma}_i),
\end{equation}
where $w_i$ are the mixture weights, and $\phi(x;\vec{\mu}_i,\vec{\Sigma}_i)$ the distribution of each individual Gaussian cluster, characterized by its mean vector $\vec{\mu}_i$ and covariance matrix $\vec{\Sigma}_i$:
\begin{equation}
\phi(x;\vec{\mu}_i,\vec{\Sigma}_i) = \frac{1}{\sqrt{(2\pi)^d|\vec{\Sigma}_i|}}e^{-\frac{1}{2}(x-\vec{\mu}_i)\vec{\Sigma}_i^{-1}(x-\vec{\mu}_i)}.
\end{equation}

\begin{figure}
\resizebox{0.99\hsize}{!}{\includegraphics{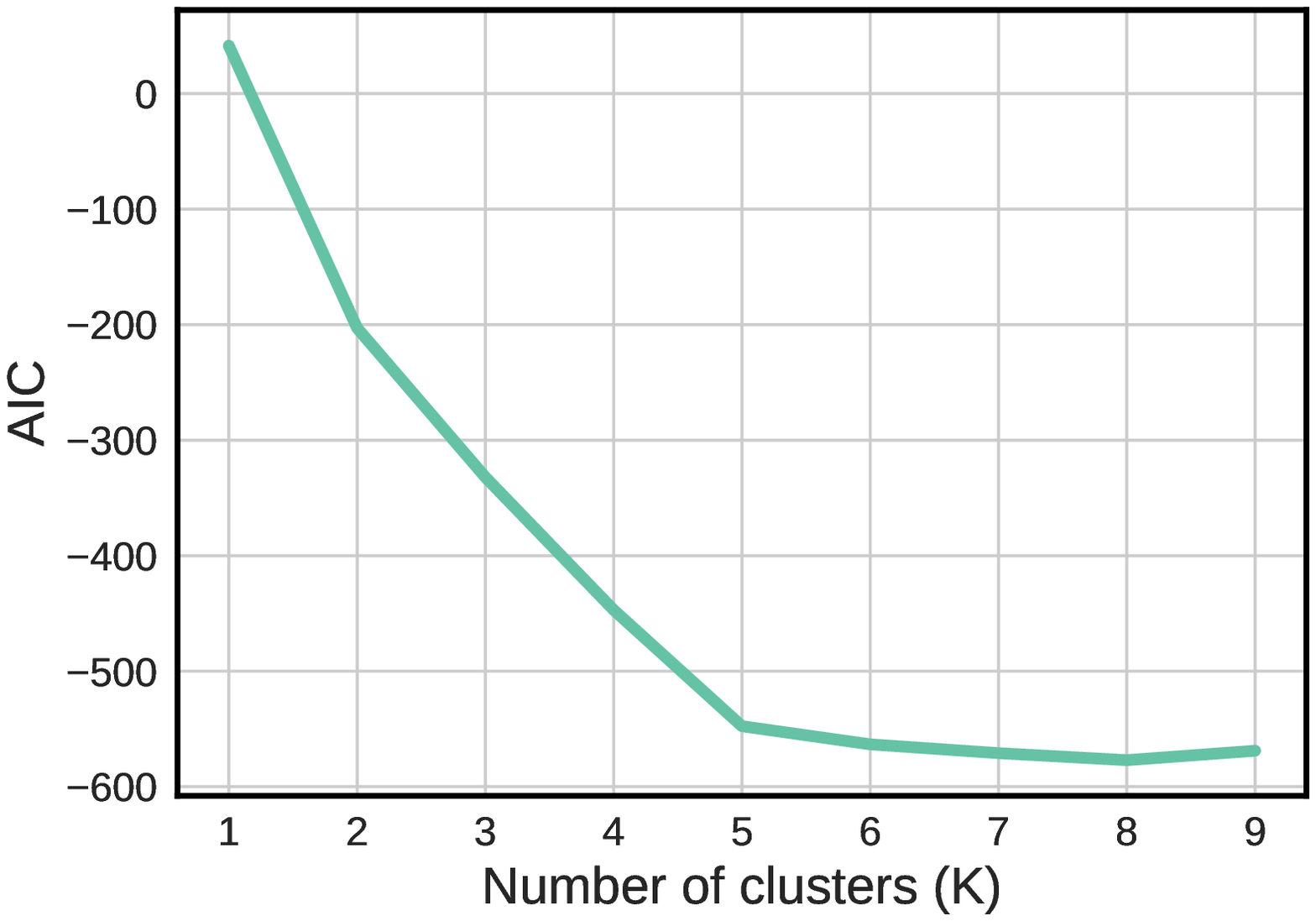}}
\resizebox{0.95\hsize}{!}{\includegraphics{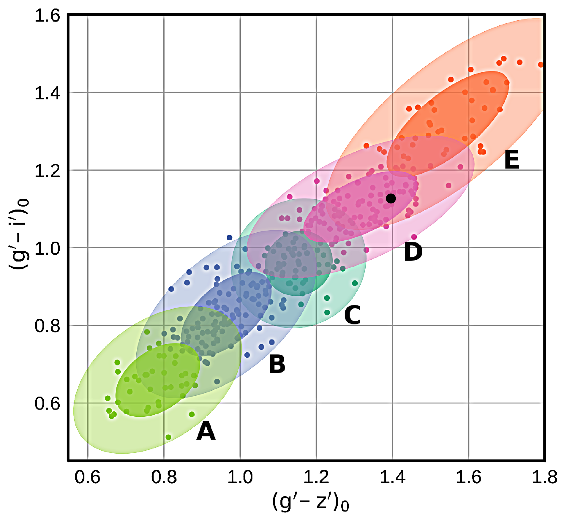}}
\caption{Upper panel: Akaike information criterion (AIC) as a function of the number of the fitted Gaussian components ($K$). Bottom panel: Colour-colour diagram background-corrected with the five-considered Gaussian components. The small and large ellipses represent 68\% and 95\% confidence intervals of the different Gaussian components, respectively. The black filled circle indicates the position of the UCD.}
\label{figure_10}
\end{figure}

In this work, we fit GMM to our dataset using the Expectation-Maximization algorithm. However, since a priori we do not know the number of groups in which the data substructure is separated, in order to determine the best number of Gaussian clusters $K$ in the CC diagram of our GC sample, we adopted the Akaike information criterion \citep[AIC;][]{akaike74}. This parameter is a measure of the relative quality of a statistical model (sum of Gaussian components) for a dataset. In this case, we obtained different AIC values for a range of $K$ components (from 1 to 9). Subsequently, we use the elbow method \citep[see e.g.,][]{baron19} to select the optimal number of groups in our data set.
Figure \ref{figure_10} (upper panel) shows that the AIC solution suggests a minimum number of five groups in our sample. The bottom panel in Figure \ref{figure_10} shows the five groups obtained by GMM in the CC diagram, where the ellipses represent the 68\% and 95\% confidence interval of the fitted Gaussian components. Different tests were performed considering GC samples towards brighter magnitudes ($g'_0<23.5$, 24, 24.5 mag). In most cases, the minimum number of fitted components was five, except for the sample with $g'_0<23.5$ mag (four components) where the ``reddest'' colour component is not observed. According to GMM, the membership of the GC candidates to each group was made considering a belonging probability superior to 50\%. Table \ref{Table_5} lists the parameters obtained by GMM considering the whole GC sample.

\begin{table}
\caption{Parameters obtained by GMM in the fitting of five Gaussian components on the CC diagram ($(g'-i')_0$ versus $(g'-z')_0$). Shown are the number of GC candidates associated with each group ($p_{i}$), the mixture weights ($w_{i}$), central vectors of the groups ($\vec{\mu}_{i}$), and covariance matrices ($\vec{\Sigma}_{i}$).}
\label{Table_5}
\begin{center}
\begin{tabular}{c|ccc}\hline\hline
	Parameter  &   value   &  \\
	\hline
        $p_{A}$ & 39   \\
        $p_{B}$ & 116  \\
        $p_{C}$ & 49   \\
        $p_{D}$ & 106  \\
        $p_{E}$ & 40   \\
	\hline        
	$w_{A}$ & 0.113 \\
	$w_{B}$ & 0.327 \\
	$w_{C}$ & 0.143 \\
        $w_{D}$ & 0.295 \\
        $w_{E}$ & 0.119 \\
	\hline
                        &  $(g'-z')$ ~~~ $(g'-i')$  &	\\
	$\vec{\mu}_{A}$	& $\left(\begin{array}{l l} 0.783 & 0.658  \end{array}\right)$	\\
	$\vec{\mu}_{B}$	& $\left(\begin{array}{l l} 0.963 & 0.821  \end{array}\right)$	\\
	$\vec{\mu}_{C}$	& $\left(\begin{array}{l l} 1.153 & 0.959  \end{array}\right)$	\\
    	$\vec{\mu}_{D}$	& $\left(\begin{array}{l l} 1.316 & 1.105  \end{array}\right)$	\\
    	$\vec{\mu}_{E}$	& $\left(\begin{array}{l l} 1.545 & 1.318  \end{array}\right)$	\\
	\hline
    
	$\vec{\Sigma}_{A}$  & $\left(\begin{array}{r r} 0.006  & 0.002 \\
                                                       0.002  & 0.004 \\
                                                 \end{array} \right)$ \\
        $\vec{\Sigma}_{B}$  & $\left(\begin{array}{r r} 0.007 & 0.003 \\
                                                       0.003 & 0.005 \\
                                                 \end{array} \right)$ \\
        $\vec{\Sigma}_{C}$  & $\left(\begin{array}{r r} 0.004 & 0.001 \\
                                                       0.001 & 0.003 \\
                                                 \end{array} \right)$ \\
        $\vec{\Sigma}_{D}$ & $\left(\begin{array}{r r} 0.011 & 0.004 \\
                                                      0.004 & 0.004 \\
                                                \end{array} \right)$ \\
        $\vec{\Sigma}_{E}$ & $\left(\begin{array}{r r} 0.012 & 0.007 \\
                                                      0.007 & 0.009 \\
                                                 \end{array} \right)$ \\
	\hline \hline
\end{tabular}
\end{center}
\end{table}

Similar to that observed in both colour histograms (upper and bottom panels in Figure \ref{figure_9}), two well-marked groups are identified, indicating the typical blue and red GC subpopulations (components B and D in Figure \ref{figure_10}). On the other hand, a third group with intermediate colours (component C), and two additional groups toward the end of the colour sample (components A and E) are detected. These last three groups could be associated with GC candidates of different ages and/or metallicities than the classic blue and red ones, or even be extensions and combinations of the latter. Figure \ref{figure_10} includes the UCD analyzed by Norris et al., which according to its integrated colours probably belongs to group D. However, it is not possible to interpret the position of this object in the colour planes in a straightforward manner because it is not a SSP and shows an extended star formation history \citep{norris15}. On the other hand, although the presence of multiple subpopulations has been studied in GC systems associated with early-type galaxies \citep{woodley10,blom12,caso15,escudero15,sesto16}, it should be considered the possibility that the use of Gaussian components for this type of analysis may not be unique. Therefore, the decomposition used in this work should be considered with caution.

\begin{figure*}
\centering
\resizebox{0.99\hsize}{!}{\includegraphics{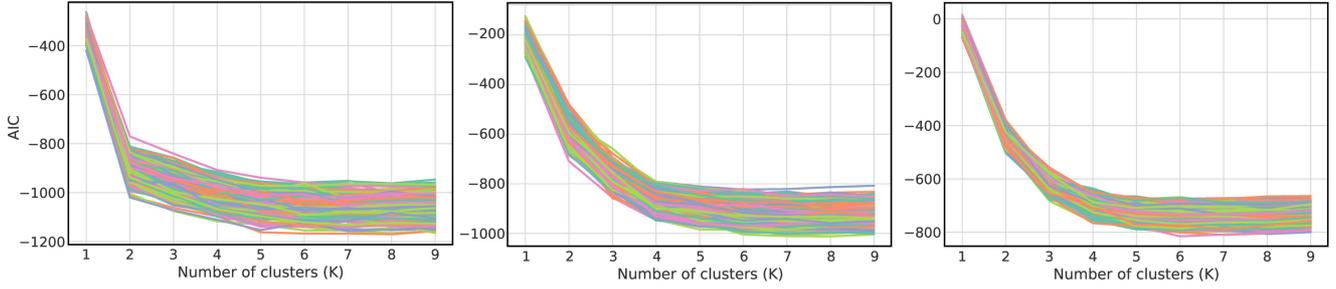}}
\caption{Akaike information criterion (AIC) as a function of the number of the fitted Gaussian components ($K$) obtained for the 500 experiments performed considering the simulated case of 2 groups of objects (left panel), 4 groups (central panel) and considering a sample of objects following a uniform distribution in colours $(g'-z')$ and $(g'-i')$ (right panel).}
\label{figure_10c}
\end{figure*}

As additional tests, we created hundreds of simulated distributions with a number of objects similar to the observed GC sample, in order to reproduce the different groups identified in the colour plane of NGC\,4546. Additionally, we add 12\% of background contamination in these samples. Initially, we simulate 2 groups of objects with mean colours and dispersions similar to the blue and red GC subpopulation of NGC\,4546, and we ran our AIC and GMM analysis to check if they were recovered. We repeat this procedure 500 times. 
Similarly, we conducted experiments considering 3 and 4 simulated groups representing the four reddest groups of NGC\,4546. Left and central panel in Figure \ref{figure_10c} shows the 500 experiments performed for the simulated cases of 2 and 4 groups, respectively.
In all cases, we obtained that the method used in this work was able to efficiently recover the simulated distributions in more than 90\% of the cases. Interestingly, we found that the inclusion of polluting background objects in the experiments makes the number of ideal components go from four to five. An inspection of this fifth component shows that it corresponds in colour to the ``A'' component previously identified in our sample of GC candidates. As the last test, we generated random samples of objects following a homogeneous distribution in the colours $(g'-z')$ and $(g'-i')$ and we ran AIC again (right panel in Figure \ref{figure_10c}). In the latter case, when quantifying the optimal number of groups obtained in each experiment, AIC estimated 4 groups in 42\% of the cases, 5 groups in 38\% and 6 groups in the remaining 20\%. Although the central panel (simulation with 4 groups) and right panel (simulation with uniform sample) of Figure \ref{figure_10c} seem to be similar at first glance, according to the values mentioned above, these results indicate that AIC cannot accurately discern the optimal number of groups for the latter case.

%============================================================================
%============================================================================

\subsection{Spatial Distribution}
\label{sec:espacial}

It is known that ``blue'' and ``red'' GCs have different 2D spatial distribution in the sense that blue GCs have more extended and less concentrated distribution than the red GCs \citep[e.g.,][]{usher13,harris17,escudero18}. Analyzing the spatial concentration toward the galaxy of the different samples of GCs candidates can help us to evaluate the reliability of our selection criteria. Additionally, merger events and/or interactions between galaxies can originate new clusters, and also affect the spatial distribution of the stellar populations as well as native GC subpopulations \citep{bonfini12,sesto18}. In this sense, the spatial anisotropy observed in some GC systems associated with early-type galaxies has been used to investigate the assembly history of the latter \citep[e.g.,][]{dabrusco14}. 

\begin{figure}
\centering
\resizebox{0.99\hsize}{!}{\includegraphics{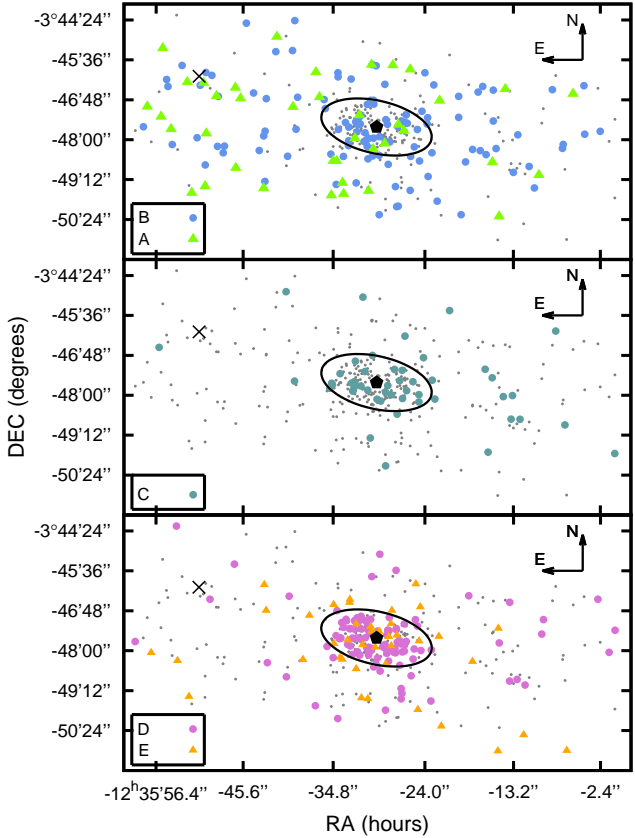}}
\caption{Projected spatial distribution of the different GC groups in NGC\,4546 assigned by GMM. Upper panel: light green triangles and blue circles represent groups A and B, respectively. Middle panel: green filled circles indicate group C. Bottom panel: magenta circles and orange triangles represent groups D and E, respectively. In all panels, the black pentagon and cross indicate the galactic centre of NGC\,4546 and the companion galaxy CGCG\,014-074, respectively. Grey dots represent the entire GC sample associated with NGC\,4546. As a comparison, the black ellipse represents the orientation and elongation of the galaxy at 5\,$R_\mathrm{eff}$ (colour figure in the online version).}
\label{figure_11}
\end{figure}

In Figure \ref{figure_11} we show the projected spatial distribution of each group of GC candidates assigned by GMM (see Section \ref{sec:histo}). As can be seen, the blue GC candidates (group B; upper panel) show an extended spatial distribution over the whole GMOS mosaic, with a slight concentration towards NGC\,4546. In contrast, the red GC subpopulation (group D; bottom panel) shows a high concentration towards the galactic centre, with approximately 72\% of its candidates located within 1.5 arcmin (6 kpc) of galactocentric radius. Similar behaviour is shown by the intermediate GC candidates (group C; middle panel). On the other hand, when observing the spatial distribution of GC candidates in the field of our mosaic containing the neighbouring galaxy CGCG\,014-074 (see Figure \ref{figure_11}), there is no significant concentration of objects around it.

Finally, GC candidates associated by GMM to groups with bluer and redder colours (groups A and E) seem to share similarities in its spatial distribution regarding the classic blue and red GC subpopulations (groups B and D), respectively. In particular, it can be seen that the objects belonging to group A do not show a significant spatial concentration on the GMOS mosaic (upper panel in Figure \ref{figure_11}). In addition, they seem to show a wide brightness range ($20.5<g'_0<24.5$ mag) when observing their location in the CM diagram ($(g'-i')_0\sim0.6$ mag; Figure \ref{figure_8}). In this context, since the galaxy does not show regions of recent star formation on its colour map (see Figure \ref{figure_5}), the presence of very blue objects with the aforementioned features would suggest that group A is probably composed of MW stars and some bona fide GCs.

\begin{figure}
\centering
\resizebox{0.95\hsize}{!}{\includegraphics{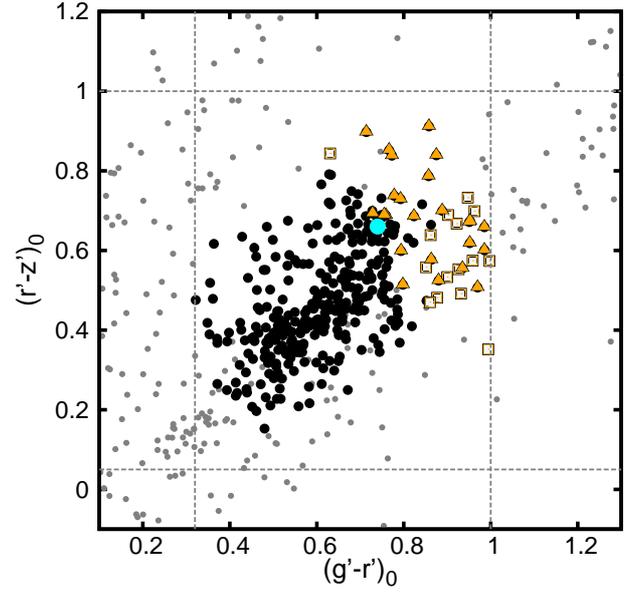}}
\caption{Colour-colour diagram with the final sample of GC candidates of NGC\,4546 (black dots), the unresolved sources detected in our GMOS mosaic (grey dots), and the GC candidates belonging to the group E according to GMM (orange symbols). Orange filled triangles and open squares indicate those objects located at $R_\mathrm{gal}<1.5$ and $R_\mathrm{gal}>1.5$ arcmin, respectively. Cyan circle indicates the UCD. The dashed grey lines show the colour ranges considered for the selection of GC candidates (colour figure in the online version).}
\label{figure_12}
\end{figure}

In the opposite case, the spatial distribution of the reddest objects (group E) shows a detectable concentration around the galactic centre. A very simple exercise shows that around 60 percent of this subsample is located inside $R_\mathrm{gal} \sim 1.5$ arcmin. However, Figure \ref{figure_8} shows that in this case, the use of a simple selection criterion such as cuts in different colours leaves GC candidates which are located on the stellar sequence of the MW. In this point, we could adopt a more restrictive cut in colours, but doing it we could reject some non-classic but genuine GCs. Therefore, in an alternative way, we analyze the spatial distribution of these candidates by splitting group E into two samples considering objects located at $R_\mathrm{gal}<1.5$ and $R_\mathrm{gal}>1.5$ arcmin (6 kpc). 
Figure \ref{figure_12} shows with orange filled triangles and open squares the location of these two subsamples in the CC diagram, respectively. As seen in the figure, objects with $R_\mathrm{gal}>1.5$ arcmin (open squares) are mostly located over the region of the stellar sequence, possibly associated with stars of the MW, while objects closer to the galaxy ($R_\mathrm{gal}<1.5$ arcmin; orange filled triangles) have a different behaviour, showing wide ranges in ($r'-z'$) colour. Therefore, the possibility that these objects are an extension of the classic red clusters and not a different subpopulation cannot be ruled out.

\begin{figure}
\resizebox{0.99\hsize}{!}{\includegraphics{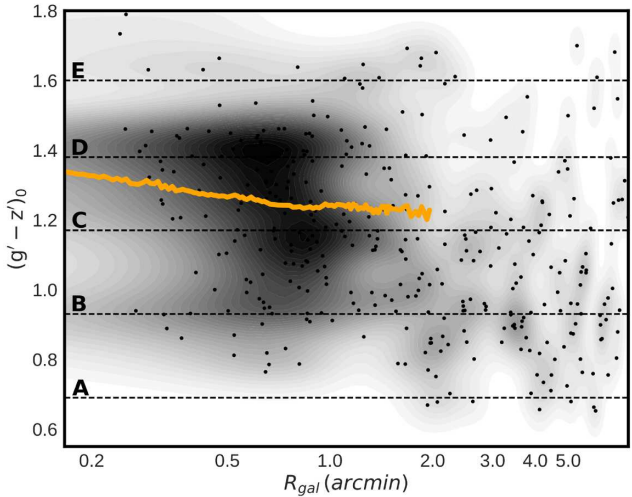}}
\caption{Smoothed colour index $(g'-z')_0$ versus galactocentric radius considering the whole sample of GC candidates (black dots). The orange solid line shows the colour profile $(g'-z')_0$ of the galaxy. The black horizontal dashed lines indicate the mean colour $(g'-z')_0$ assigned by GMM to the five groups.}
\label{figure_13}
\end{figure}

The previous qualitative analysis can be also seen in the smoothed colour index $(g'-z')_0$ diagram as a function of the projected galactocentric radius ($R_\mathrm{gal}$). As seen in Figure \ref{figure_13}, the blue GC subpopulation (group B) is extended over the GMOS mosaic ($R_\mathrm{gal}\gtrsim0.25$ arcmin; $\gtrsim 1$ kpc), while the groups associated with the intermediate, red and reddest GC candidates (groups C, D and E) show a concentrated spatial distribution mainly between $0.25<R_\mathrm{gal}<2.4$ arcmin ($1-9.6$ kpc). On the other hand, the objects belonging to the bluest group (A), shows marginal clumps between $\sim$0.7 and 6 arcmin ($2.8-24$ kpc). In order to assess whether this last group is still detected towards smaller galactocentric radii, we performed a test considering those objects located at $R_\mathrm{gal}<1.5$ arcmin. In this case, Akaike's solution suggests that at least 4 groups represent this sample, where groups B, C, D and E are identified again. Therefore, this reinforces the idea that group ``A'' could be a mixture of objects between some bona fide GCs of NGC\,4546 and MW stars.

%%%%%%%%%%%%%%%%%%%%%%%%%%%%%%%%%%%%%%%%%%%%%%%%%%%%

\subsection{Integrated Colours and their Comparison with SSP Models}
\label{sec:ssp}

Due to the well known ``age-metallicity degeneracy'' effect in optical colors, it is not possible to obtain unique ages and metallicities estimations from them. However, we can use them to check if some SSP models can reproduce the observed mean values for the different GC subpopulations identified in Section \ref{sec:histo}, and thus obtain any useful indication regarding the composition of these subsamples.

\begin{figure}
\resizebox{0.99\hsize}{!}{\includegraphics{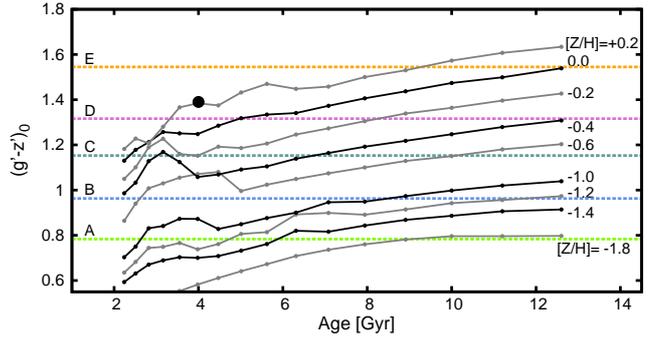}}
\caption{Mean colours $(g'-z')_0$ (horizontal dashed lines) obtained by GMM for each group compared with single stellar population models of \citet{bressan12}. The black and gray lines indicate the different metallicities used ($\mathrm{[Z/H]}=-1.8, -1.4, -1.2, -1.0, -0.6, -0.4, -0.2, 0.0, +0.2$ dex from bottom to top). Black filled circle shows the location of the UCD.}
\label{figure_14}
\end{figure}

In order to do that, we compared the mean colour $(g'-z')_0$ obtained by GMM for each group with the SSP models of \citet{bressan12} (Figure \ref{figure_14}). As can be seen in \citet{sesto16}, the photometric predictions of these models are in good agreement with observed integrated GC colours in the Sloan photometric system. We consider different metallicity values between $\mathrm{[Z/H]}=-1.8$ to $+0.2$ dex to cover the typical values for blue and red GC subpopulations \citep[e.g.,][]{usher12}. Assuming old ages between 8 and 12 Gyr for these two groups, the mean metallicity values obtained for them result in $\mathrm{[Z/H]}\sim-1.2$ and $\sim-0.3$ dex, respectively. In the case of the intermediate group in our sample (group C), we see that similar metallicities to the red GC candidates (i.e. group D, with $\mathrm{[Z/H]}\sim-0.4$ to $-0.2$ dex) in the age range of $\sim3-7$ Gyr, or metallicity values around $\sim-0.6$ dex for an age of $\sim$10 Gyr could reproduce the observed colours. Although values of $\mathrm{[Z/H]}<-0.4$ dex, and therefore ages $<3$ Gyr, could be fitted to the intermediate candidates, these clusters are not significantly bright objects in our sample as seen in other galaxies with recent star formation, such as NGC\,1316 \citep{sesto18}. Therefore, this last option does not seem to be the case in NGC\,4546. On the other hand, the bluer and redder groups defined by GMM (groups A and E) would be associated with objects with metallicities $\mathrm{[Z/H]}\lesssim-1.0$ dex ($>$2 Gyr) and $\mathrm{[Z/H]}>+0.0$ dex ($\gtrsim7$ Gyr), respectively. Figure \ref{figure_14} also includes the position of the UCD analyzed by \citet{norris15} showing that its light is well represented by an SSP of around 4 Gyr old and a supra solar metallicity. This is consistent with the values obtained by these authors through the analysis of Lick indices.

%%%%%%%%%%%%%%%%%%%%%%%%%%%%%%%%%%%%%%%%%%%%%%%%%%%%

\subsection{Radial and Azimuthal Distribution}
\label{sec:azimutal}
With the aim of quantifying the radial and azimuthal distributions of the GC candidates in NGC\,4546, we consider the following guidelines.
Initially, we obtained the projected surface density profiles for the whole sample of clusters, as well as for the different subpopulations identified in Section \ref{sec:histo}. For this, we constructed one-dimensional radial distributions counting objects in concentric circular rings with steps between $\Delta\,log\,r=0.08$ and 0.12, according to the sample analyzed. Each ring was corrected for contamination and effective area. This last correction was applied in those cases where the ring exceeded the physical limit of the GMOS mosaic. The uncertainties in the surface density profiles are given by the Poisson statistic.

In order to describe the radial distribution of the different GC groups, we fit several functions to our profiles. A power-law ($log\,\sigma_{GC}=a+b\,log(r)$), de Vaucouleurs ($log\,\sigma_{GC}=a+b\,r^{1/4}$), and S\'ersic profiles (i.e. $r^{1/n}$) were used, with the aim to assess which of these scaling laws best represent each profile, and also to obtain some relevant parameters. 

\begin{figure}
\centering
\resizebox{0.95\hsize}{!}{\includegraphics{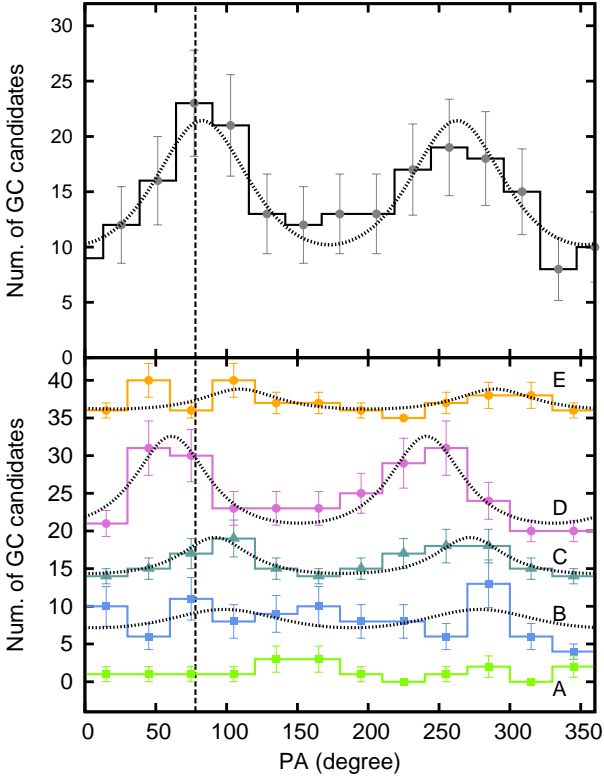}}
\caption{Azimuthal distribution for the whole sample of GC candidates (upper panel), and for the different GC groups assigned by GMM (bottom panel) within the circular ring $0.36<R_\mathrm{gal}<2.4$ arcmin ($1.5<R_\mathrm{gal}<9.8$ kpc). The different colours used in the histograms are the same as those assigned for each GC subpopulation as Figure \ref{figure_11}. In order to avoid overlapping, the histograms corresponding to blue, intermediate, red and reddest GC candidates were vertically shifted adding 3,13,18 and 35 to the counts, respectively. Dotted lines in both panels show the fits obtained using expression \ref{eq:acim}. The vertical dashed line indicates the $PA$ of the galaxy.}
\label{figure_15}
\end{figure}

However, as seen in Figure \ref{figure_11}, some GC subpopulations show a significant elongation in their spatial distribution, mainly the intermediate and red subpopulations (groups C and D). Therefore, in order to obtain appropriate density profiles for each GC sample according to its orientation and spatial elongation, and subsequently, to achieve better fits of the aforementioned functions, we decided to study the azimuthal distribution of the different GC groups. 
For this, we constructed histograms of the azimuthal distribution for the whole GC sample, as well as for each GC group defined by GMM (Section \ref{sec:histo}). We count objects in different wedges within a circular ring of $0.36<R_\mathrm{gal}<2.4$ arcmin ($1.5<R_\mathrm{gal}<9.8$ kpc) centred on the galaxy. The inner radius was chosen to avoid the incompleteness effect in the region near the galactic centre. On the other hand, the outer radius was chosen to cover the largest area within our GMOS mosaic, and thus avoid the correction by areal incompleteness in the counts. Figure \ref{figure_15} shows the different histograms of this analysis. 

Subsequently, we determined the ellipticity ($\epsilon$) and position angle ($PA$) of each sample using the expression of \citet{mclaughlin94} given by:

\begin{equation}\label{eq:acim}
  \sigma(R,\theta)=kR^{-\alpha}[cos^2(\theta-PA)+(1-\epsilon)^{-2}sin^2(\theta-PA)]^{-\alpha/2},
\end{equation}

\noindent where $\sigma(R,\theta)$ is the number of GC candidates, $k$ the normalization constant, $\theta$ is the $PA$ measured counterclockwise from the north, and $\alpha$ the value of the power-law exponent in the surface density fit. During the fit process, the free variation of the different parameters was allowed. Table \ref{Table_6} lists the shape parameters obtained from the azimuthal analysis and those of the stellar light distribution of the galaxy (see Section \ref{sec:photo}). 

\begin{table}
\centering
%\scriptsize
\begin{tabular}{lcc}
\multicolumn{3}{c}{}\\
\toprule
\toprule
\multicolumn{1}{c}{\textbf{Population}} &
\multicolumn{1}{c}{\textbf{$\epsilon$}} &
\multicolumn{1}{c}{\textbf{$PA$\,(degrees)}} \\
\hline
All          & 0.30$\pm$0.04  & 83$\pm$4   \\
Blue (B)     & 0.23$\pm$0.18  & 99$\pm$25  \\
Interm. (C)  & 0.50$\pm$0.06  & 92$\pm$3   \\
Red     (D)  & 0.41$\pm$0.05  & 61$\pm$6   \\
+Red    (E)  & 0.46$\pm$0.20  & 109$\pm$15 \\
Gal. light   & 0.50$\pm$0.01  & 79$\pm$2   \\
\bottomrule
\end{tabular}
\caption{Ellipticity ($\epsilon$) and position angle ($PA$) values for the GC system and for each subpopulation in NGC\,4546. In addition, the mean values of these parameters obtained for the galaxy light in the range $0.2<r_\mathrm{eq}<2$ arcmin (see Section \ref{sec:ellipse}) are listed.}
\label{Table_6}
\end{table}

We find for the whole GC system an ellipticity of $\epsilon=0.30\pm0.04$ while the galaxy light has a higher value of 0.50$\pm$0.01. This difference probably results from a combination between the azimuthal values obtained for the blue candidates (group B), which do not show a significant ellipticity and orientation and that is reflected in their high uncertainties values, and in the relatively high values of the intermediate ($\epsilon=0.50\pm0.06$; group C) and red GC subpopulations ($\epsilon=0.41\pm0.05$; group D).
However, the position angle between the GC system and the light of NGC\,4546 are in agreement within the errors ($PA=83\degree\pm4\degree$ and $79\degree\pm2\degree$, respectively).
On the other hand, as listed in Table \ref{Table_6}, intermediate and red GC subpopulations (group C and D) show similar values in their ellipticity as compared to the starlight of NGC\,4546, but with slightly different orientations between them. In the case of the reddest GC candidates (group E), even with a low number of objects in the considered region, it was possible to determine $PA$ and an ellipticity, the latter being similar to those of C and D subpopulations.

\begin{figure*}
\centering
\resizebox{0.95\hsize}{!}{\includegraphics{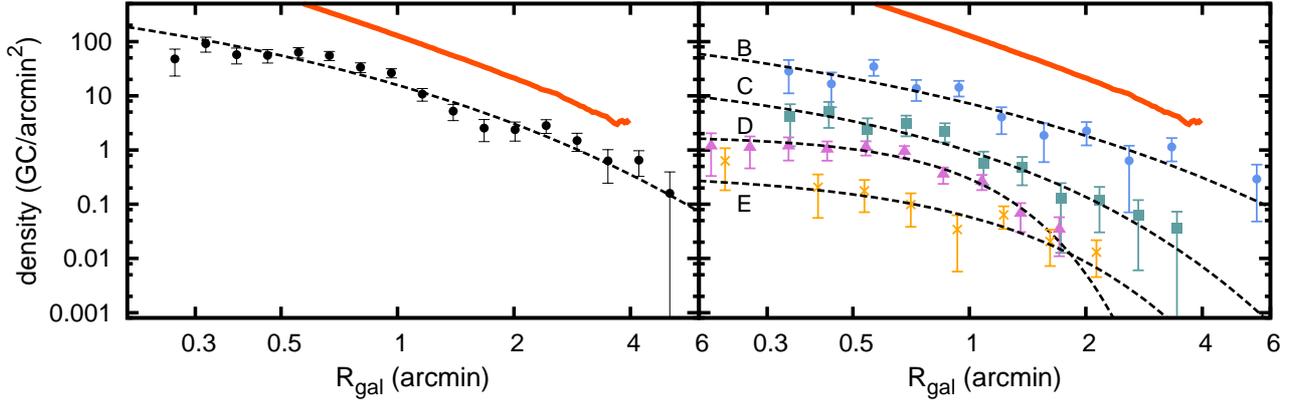}}
\caption{Density profiles corrected by contamination and areal completeness for the total GC sample (black filled circles in left panel) and for each GC subpopulations in NGC\,4546 (right panel). Blue circles, green squares, magenta triangles and orange crosses in right panel indicate the blue, intermediate, red and reddest GC subpopulations (groups B, C, D and E), respectively. These profiles were shifted along the y-axis to avoid overlapping. The dashed black line indicates the fit obtained using the S\'ersic function. The solid orange line indicates the galaxy starlight profile (colour figure in the online version).}
\label{figure_16}
\end{figure*}

With the information obtained from this analysis, we reconstructed the density profiles for the whole sample as well as for each group of GCs. The case of the intermediate, red and reddest GC subpopulations (groups C, D and E), we considered concentric elliptical rings for their density profiles according to its position angle and ellipticity values. This was also considered for the whole GC sample. On the other hand, we kept circular rings for the blue GC subpopulation (group B).
Figure \ref{figure_16} shows the different background corrected density profiles along with the galaxy starlight profile (see Section \ref{sec:photo}) as a function of the galactocentric radius. As seen in this figure, the density profiles of groups B, C, D and E show a decreasing behaviour as we move away from the galactic centre. In contrast, the objects belonging to group A do not show such behaviour, therefore their density profile was not included in the figure.

Afterwards, we fit the power-law, de Vaucouleurs, and S\'ersic functions excluding those points centrally located ($r<0.3$ arcmin) in order to avoid incompleteness effects. Tables \ref{Table_7} and \ref{Table_8} list the parameters and reduced chi-squared values obtained in the fits. The S\'ersic profile for the reddest objects (group E) showed a high uncertainty in its parameters, probably due to the low counts in its density profile. Therefore, the values obtained by keeping free all the parameters during the fit for this group were not included in Table \ref{Table_8}. However, we found that an exponential profile (i.e. $n=1$) gives a reasonable fit this group. Therefore, we include the obtained $R_\mathrm{eff}$ and reduced chi-squared values for this fit in Table \ref{Table_8}.

\begin{table}
\centering
%\scriptsize
\begin{tabular}{lcccc}
\multicolumn{5}{c}{}\\
\toprule
\toprule
\multicolumn{1}{c}{\textbf{Population}} &
\multicolumn{1}{c}{\textbf{Scal. law}} &
\multicolumn{1}{c}{\textbf{Slope ({\it b})}} &
\multicolumn{1}{c}{\textbf{Zero point ({\it a})}} &
\multicolumn{1}{c}{\textbf{$\chi^2_\nu$}} \\
\hline
All      & Power     & -2.48$\pm$0.13  & ~2.11$\pm$0.25  & 1.3  \\
         & de Vauc.  & -1.41$\pm$0.08  & ~1.65$\pm$0.25  & 1.5  \\
\hline
Blue (B) & Power     & -1.76$\pm$0.21  & ~0.27$\pm$0.41  & 1.6  \\
         & de Vauc.  & -0.94$\pm$0.12  & -0.21$\pm$0.39  & 1.9  \\
\hline
Interm. (C) & Power     & -2.18$\pm$0.21  & ~0.79$\pm$0.41  & 1.2  \\
            & de Vauc.  & -1.58$\pm$0.14  & ~1.32$\pm$0.42  & 0.7  \\
\hline
Red (D)  & Power     & -2.94$\pm$0.39  & ~2.45$\pm$0.66  & 1.2  \\
         & de Vauc.  & -1.89$\pm$0.22  & ~2.52$\pm$0.59  & 0.9  \\
\hline
+Red (E) & Power     & -1.87$\pm$0.26  & ~0.04$\pm$0.49  & 0.6  \\
         & de Vauc.  & -1.09$\pm$0.14  & -0.21$\pm$0.42  & 0.5  \\    
\bottomrule
\end{tabular}
\caption{
Slope ({\it b}) and zero point ({\it a}) values obtained in the fit of the density profiles considering a power-law and de Vaucouleurs functions. The last column indicates the values of reduced chi-square.}
\label{Table_7}
\end{table}

\begin{table}
\centering
%\scriptsize
\begin{tabular}{lccc}
\multicolumn{4}{c}{}\\
\toprule
\toprule
\multicolumn{1}{c}{\textbf{Population}} &
\multicolumn{1}{c}{\textbf{$n$}} &
\multicolumn{1}{c}{\textbf{$R_\mathrm{eff}$ ($''$)}} &
\multicolumn{1}{c}{\textbf{$\chi^2_\nu$}} \\
\hline
All           & 2.5$\pm$1.3  & 46.7$\pm$9.5  & 1.8 \\
Blue (B)      & 1.2$\pm$0.7  & 53.4$\pm$9.5  & 0.8 \\
Interm. (C)   & 1.6$\pm$0.8  & 41.5$\pm$8.4  & 0.5 \\
Red (D)       & 0.6$\pm$0.2  & 38.3$\pm$2.9  & 0.6 \\
+Red (E)      & 1.0          & 52.4$\pm$13.1 & 0.6 \\
\bottomrule
\end{tabular}
\caption{S\'ersic parameters obtained in the fit of the density profile for the whole GC sample, as well as for blue, intermediate, red and reddest GC subpopulations (groups B, C, D and E, respectively). The last column indicates the values of reduced chi-square.}
\label{Table_8}
\end{table}

This analysis indicates that blue GC subpopulation (group B) has a larger effective radius than the intermediate and red GC subpopulations (groups C and D), while $R_\mathrm{eff}$ values of the latter pair are similar. In this sense, the S\'ersic function provides a better fit for the blue candidates, while for the intermediate and red GC candidates, the respective power-law and de Vaucouleurs functions achieve a better fit. As seen in Figure \ref{figure_16} and listed in Tables \ref{Table_7} and \ref{Table_8}, red GC candidates present a higher density and concentration towards NGC\,4546, showing a low value of the S\'ersic index ($n<1$) that gives its truncated shape.

In addition, Table \ref{Table_8} indicates that the S\'ersic profile for the whole sample gives a fit worst than in the case of power-law and de Vaucouleurs functions. This it seems to be related to a slight over-density of objects around $r_\mathrm{eq}\sim0.66$ arcmin (2.7 kpc) which make a single S\'ersic law not suitable for describing that profile. Interesting, this position coincides with the galactocentric radius where the highest concentration of red and intermediate GC candidates is found (see Figure \ref{figure_13}).

%============================================================================
%============================================================================

\subsection{Luminosity Function and Total Number of GCs} 
\label{sec:lumin}
In order to determine the total GC population in NGC\,4546 and also to estimate a new value of the distance to the galaxy, we obtained the luminosity function of the globular clusters (GCLF). To do this, we select the unresolved objects in our initial photometric catalogue by performing the same colour cuts mentioned in Section \ref{sec:color}, but without restricting the magnitudes in the $g'$ band. We counted objects in bins of 0.35 mag in the range $17<g'_0<27$ mag, correcting by completeness and contamination using the analysis carried out in Sections \ref{sec:test} and \ref{sec:comp}, respectively. 

\begin{figure}
\centering
\resizebox{0.99\hsize}{!}{\includegraphics{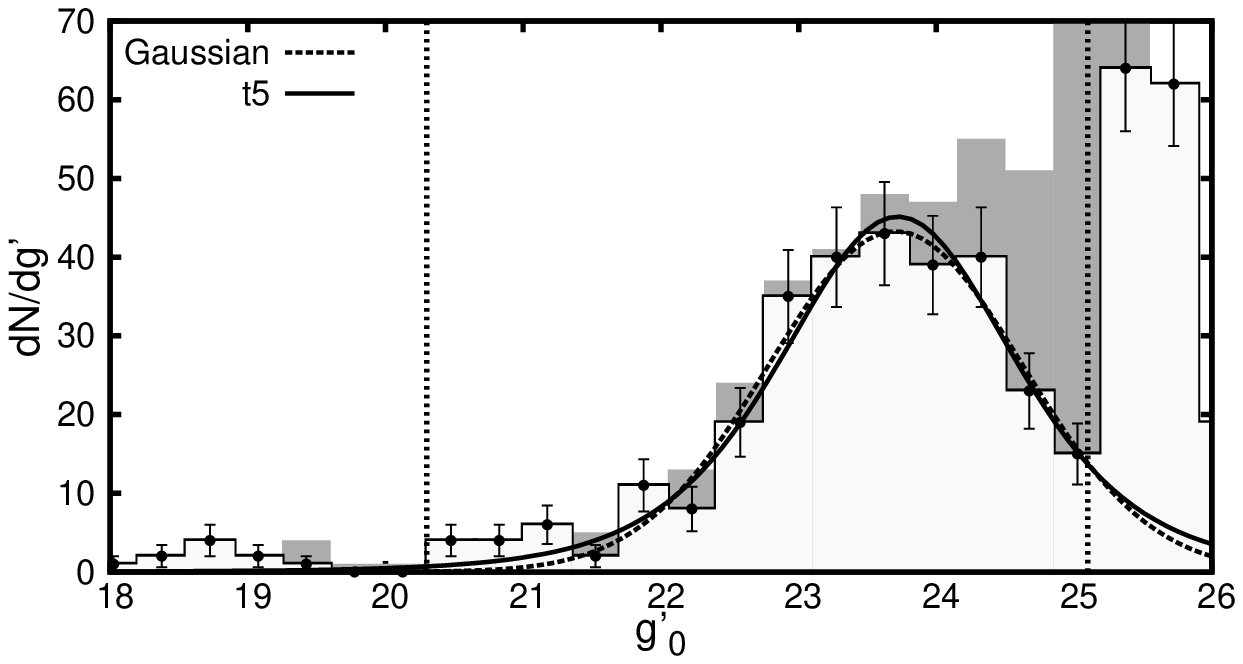}}
\resizebox{0.99\hsize}{!}{\includegraphics{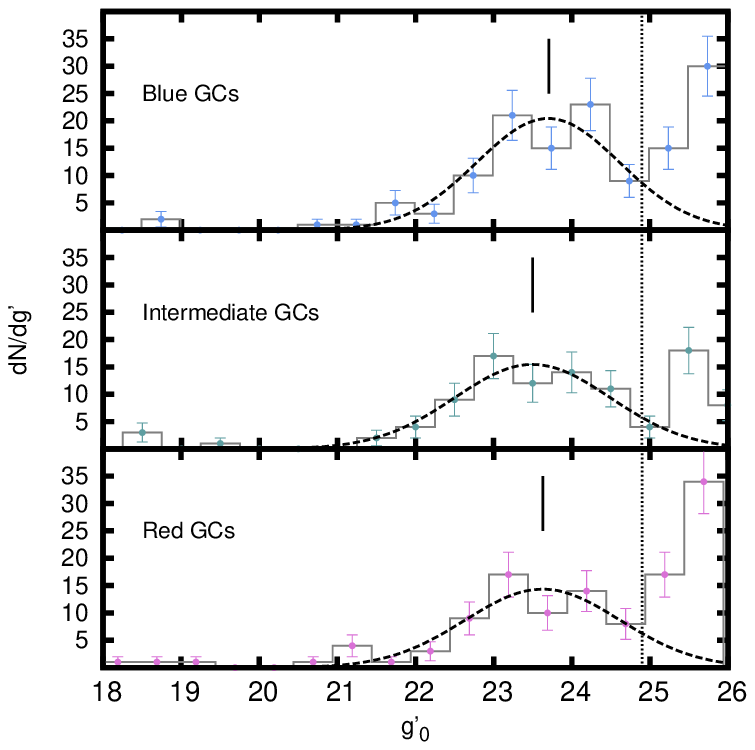}}
\caption{Upper panel: raw histogram (shaded histogram), and background and completeness corrected histogram (black solid line) for the brightness distribution of the GC system of NGC\,4546. Vertical dotted lines indicate the magnitude region used for the fit of the Gaussian (black dashed line) and t5 (black solid line) functions. Bottom panel: corrected GCLFs for the blue, intermediate and red GC subpopulations (groups B, C and D). Vertical dotted and solid lines indicate the magnitude limit used for the fits, and the TO positions, respectively.}
\label{figure_17}
\end{figure}

Figure \ref{figure_17} (upper panel) shows the raw luminosity distribution (grey line and scratched histogram), and the completeness and background corrected histogram (black line and white histogram) for the GC candidates of NGC\,4546. In order to model the GCLF and obtain the position of the turnover ($TO$), we fit the Gaussian and t5 functions down to the 80\% completeness limit ($g'_0=25.2$ mag). As seen in the figure, from this value an increase in the count of objects is observed, possibly as a consequence of an imperfect correction of the completeness and/or contamination of the sample. The values obtained were $TO_G=23.70\pm0.08$, $\sigma_G=0.92\pm0.07$ mag and $TO_{t5}=23.72\pm0.07$, $\sigma_{t5}=0.88\pm0.07$ mag for the Gaussian and t5 function, respectively. These values are reasonably robust thanks to the depth of our photometric sample.

In this work, we adopted the average values of $TO$ and $\sigma$ of those values mentioned above, obtaining $\langle TO \rangle=23.71\pm0.1$ and $\langle \sigma \rangle=0.9\pm0.09$ mag for the GCLF of NGC\,4546. Assuming a universal value of TO of $M_{V}=-7.4$ mag \citep{jordan07} and using the equation 2 of \citet{faifer11} to transform $M_{V}$ to $M_g$, we obtained the distance modulus of $(m-M)=30.75\pm0.12$ mag ($d=14.1\pm1.0$ Mpc). This value is in excellent agreement with that previously obtained by \citet{tully13} ($(m-M)=30.73\pm0.14$ mag) using surface brightness fluctuation method. 

Based on the information obtained from the surface density profile of the whole GC sample and from the GCLF analysis, we estimate the total GC population and the specific frequency ($S_N$) of NGC\,4546.
For this, we follow the same procedure as in \citet{escudero15,escudero18}. We integrated the density profile from 0.26 arcmin up to the galactocentric radius of 12.3 arcmin, equivalent to 50 kpc. The inner radius was chosen due to incompleteness in our data towards the galactic centre. On the other hand, the value adopted for the outermost radius is appropriate to include most of the GCs associated with an intermediate-mass galaxy \citep{norris11,faifer11}. For the region $R_\mathrm{gal}<0.26$ arcmin, we considered the constant value of 93 GC/arcmin$^{2}$ for the density, which is associated with the second bin in the density profile. This gives us 370 GCs. 
According to the GCLF parameters used in this work, this value represents 95 per cent of the whole GC sample (GCs brighter than $g'_0=25.2$ mag). Then, the total population of NGC\,4546 is $N_\mathrm{tot}=390\pm60$ GCs. The error associated with the total number of clusters is mainly due to the uncertainty in the parameters obtained in the analysis of the radial distribution and the luminosity function.
 
Using $N_\mathrm{tot}$, we obtained the $S_N$ of the system defined as the total number of GCs per unit host galaxy luminosity. This parameter allows comparisons between GC systems with different star formation histories and, therefore, different mass-luminosity ratios in their stellar populations. The expression initially introduced by \citet{harris81} is defined as:
\begin{equation}\label{specific}
S_N = N_{tot}10^{0.4(M_V+15)},
\end{equation}
where $M_V$ is the absolute magnitude of the galaxy in the $V$ band. Using the previously obtained distance modulus and the total $V$ magnitude $V=10.57\pm0.01$ of the Carnegie-Irvine Galaxy Survey \citep{ho11}, the absolute magnitude of NGC\,4546 is $M_V=-20.18\pm0.12$. Then, we obtained a specific frequency of $S_N=3.3\pm0.7$. This is a high value when compared to the mean value $S_N=1.6\pm1.4$ obtained for GC systems associated with S0 galaxies from the \citet{harris13} catalogue. 

In order to investigate if the intermediate GC candidates have any influence in the $S_N$ value, we obtained the GCLF of the blue, intermediate and red GC subpopulations (groups B, C and D; see Section \ref{sec:histo}). Due to the low number of candidates in each of these groups, we considered bins of 0.5 mag to construct the GCLFs. Figure \ref{figure_17} (bottom panel) and Table \ref{Table_9} show the corrected GCLFs and the parameter values obtained from the fitted functions, respectively.

From this analysis, we estimate the total number of GCs in these three GC subpopulations, obtaining 146$\pm$25 clusters for the blue (group B), 64$\pm$10 members for the intermediate (group C), and 115$\pm$20 for the red ones (group D). Considering only the values corresponding to the blue and red GC subpopulations, we calculate again the $S_N$ for NGC\,4546, obtaining in this case, $S_N=2.2\pm0.5$. This value is more consistent with the mean value obtained for this type of galaxies.
However, even if we consider these three subsamples, the red fraction of GC candidates in NGC 4546 results to be $f_r=0.35\pm0.10$. This value is similar to those presented for galaxies with similar total stellar masses by \citet{salinas15}.

It is worth to mention that as the $S_N$ parameter depends on $M_V$, which in turn depends strongly on age, their values can be affected by age differences among stellar populations in galaxies. Therefore, we consider two quantities that are not affected by such differences, the $T_N$ parameter \citep{zepf93}, and the specific mass $S_M$ \citep{georgiev10}. $T_N$ is the number of GCs normalized to the stellar mass in units of $1\times10^9 M_\odot$, while $S_M$ is the total mass of the GC system relative to the total baryon mass of the host galaxy defined as the sum of the stellar mass and the H{\sc{i}} gas mass. For the latter, we follow the guidelines of \citet{harris17} to estimate the total mass of the GC system. Then, considering the value of stellar mass of NGC\,4546 from Table \ref{Table_1} and the H{\sc{i}} gas mass of \citet{bettoni91}, high values were obtained for both quantities, $T_N=14.4$ and $S_M=0.28$ \citep[see, e.g., Figure 6 of][]{georgiev10}. In the same way, in order to compare some of these results with values in the literature, we obtain $T_N$ for blue and red GC subpopulations: $T_\mathrm{blue}=5.4$ and $T_\mathrm{red}=4.3$, respectively. These values compared with those obtained for the sample of galaxies belonging to low-density environments presented by \citet{salinas15}, show that NGC\,4546 has a relatively rich GC system for its mass, comparable to those of early-type galaxies in clusters.

\begin{table}
\centering
%\scriptsize
\begin{tabular}{cccc}
\multicolumn{1}{c}{ } \\
\toprule
\toprule
\multicolumn{1}{c}{\textbf{Subpopulation}} &
\multicolumn{1}{c}{\textbf{function}} &
\multicolumn{1}{c}{\textbf{$TO$}} &
\multicolumn{1}{c}{\textbf{$\sigma$}} \\
\midrule
Blue         & Gauss. & $23.70\pm0.15$ & $0.91\pm0.17$ \\
             & t5     & $23.71\pm0.16$ & $0.93\pm0.19$ \\
Intermediate & Gauss. & $23.50\pm0.11$ & $0.99\pm0.12$ \\
             & t5     & $23.50\pm0.12$ & $1.00\pm0.15$ \\
Red          & Gauss. & $23.63\pm0.18$ & $0.99\pm0.19$ \\
             & t5     & $23.62\pm0.18$ & $1.00\pm0.20$ \\
\bottomrule
\end{tabular}
\caption{$TO$ and $\sigma$ values obtained from the fits of the Gaussian and t5 functions for the GCLFs of blue, intermediate and red GC subpopulations (groups B, C and D).} 
\label{Table_9}
\end{table}

%===========================================================================
%===========================================================================

\section{Summary and Conclusions}
\label{sec:summary}
In this work we present the analysis of the global properties of the GC system associated with the S0 galaxy NGC\,4546, using Gemini/GMOS data of three fields in the $g'r'i'z'$ filters. A summary of the main photometric results is presented below.

The analysis of the surface brightness distribution of the galaxy reveals considerable variations in its isophotal parameters towards the inner region ($r_\mathrm{eq}<0.2$ arcmin; $0.8$ kpc), mainly in its position angle and ellipticity. A 2D image decomposition using {\sc{galfit}} software indicates that the galaxy is well fit with four components (bulge + lenses/disks + halo). The residual image obtained from subtracting this model in our GMOS images does not reveal the presence of a possible bar structure or spiral pattern as mentioned by \citet{bettoni91}. 
In addition, extensive irregular regions of dust extending about 6 kpc along the semi-major axis of the galaxy are observed.
These structures, together with the presence of a disk of gas rotating in the opposite direction to the stellar component of the galaxy, provide evidence of a merger/interaction event in the recent past with an object of lower mass than NGC\,4546 \citep{eliche18}. In this context, the presence of the young NGC\,4546-UCD1 with an age of $\sim4$ Gyr and with a rotation direction similar to the gas distribution, would be related to the aforementioned interaction event. However, a possible previous interaction with the small companion S0 galaxy (CGCG\,014-074) located at a projected distance of $\sim22$ kpc cannot be ruled out either.

Regarding the study of the GC system of NGC\,4546, we measured the photometry of 350 GC candidates obtaining the integrated colour distributions $(g'-i')_0$ and $(g'-z')_0$, their density profiles and azimuthal distributions. We estimated the distance modulus of the galaxy through its GCLF, resulting in $(m-M)=30.75\pm0.12$ mag. In addition, assuming a maximum length of 50 kpc for the GC system of NGC\,4546 we obtained a total population of $N_\mathrm{tot}=390\pm60$ GCs. This total number of GC candidates is significantly higher than the previously estimated value of \citet{zaritsky15}, and it corresponds to a relatively high specific frequency for this type of galaxies of $S_N=3.3\pm0.7$. While this is not as extreme as the value of $S_N=10$ obtained for NGC\,6861 \citep{escudero15}, another low-density environment lenticular galaxy, a comparison with the sample compiled in \citet{salinas15} shows that NGC\,4546 and NGC\,6861 support the view that galaxies in such environment can have $S_N$ values in a similar range to that of cluster early-type galaxies. 

We studied the $(g'-i')_0$ and $(g'-z')_0$ GCs colour distribution throughout the fit of Gaussian functions to the histograms. Our analysis shows that the bimodal case results in a GC system strongly dominated by the red subpopulation. An interpretation of this observation could be that NGC\,4546 has a relatively quiescent merger history. This is because it is thought that many, if not most, metal-poor GCs are gained from satellite accretion \citep[see e.g.,][]{forbes10,choksi18}. This scenario is supported by the case of NGC\,1277, a galaxy that is strongly suspected of having undergone very few mergers because it completely lacks blue GCs \citep[a so-called ``red nugget galaxy'';][]{beasley18}. However, NGC\,4546 is an intermediate luminosity galaxy displaying a relatively high $S_N$ (and $T_\mathrm{blue}$) in a low-density environment. In this context, that interpretation may not be suitable as it would mean that with a normal history of mergers, NGC\,4546 would have had an even higher $S_N$.

Therefore, we carried out a statistical analysis of our GCs sample in the colour-colour plane which showed different components, indicating at least three or more distinct GC groups. The classic blue and red subpopulations are observed, a group with intermediate colours to those mentioned above, and two groups located towards the blue and red ends of the colour distributions. However, a follow-up spectroscopic study will be necessary to confirm whether these groups are distinct in their stellar populations or dynamics. In this work, as seen in other early-type galaxies, the blue GC subpopulation presents a flatter spatial distribution than the red one, which has a high concentration degree towards the galaxy centre. Noticeably about 72\% of the red candidates in our sample are located within 6 kpc of galactocentric radius and a fit of a S\'ersic law gave an index $n<1$. This is clearly reflected in the density profile obtained for the whole sample where a slight overdensity is seen at $R_\mathrm{gal}\sim1$ arcmin. 
In the case of the intermediate GC subpopulation, its density profile presents a slope value between those obtained for the blue and red group, which indicates features shared with them. In this context, the analysis of the azimuthal distribution shows that intermediate and red GC candidates have similar values in their ellipticities with respect to the stellar light distribution of NGC\,4546, while blue candidates do not show a significative ellipticity. These results would indicate a possible common origin between the red GCs and the stellar component of the galaxy \citep{forte14}, and possibly also that of the intermediate GCs if they were a different group.
However, the value of the $PA$ obtained for the red subpopulation differs significantly from the values obtained for the other group of candidates and for the stellar light of the galaxy. This misalignment could be another indication of the violent past of this galaxy.

On the other hand, the different features shown by the blue GC subpopulation would indicate that this group would be a mixture between clusters originating {\it in situ} as well as accreted clusters from low-mass dwarf galaxies during the mass assembly phase of the galaxy. As NGC\,4546 is a low mass object in a very low-density environment, it might be expected that it shows evidence of a deficit of blue GCs due to a low accretion efficiency. However, this is not the case and therefore it would seem to indicate that the blue part of this GC system was assembled in a similar way to other galaxies. 

Regarding the bluest GC objects, they do not show a particular characteristic or concentration around the galaxy. The presence of these objects with very blue colours distributed over our GMOS mosaic, possibly indicates that this group is composed of some bona fide GCs but mainly by MW stars. Only spectroscopic follow up can confirm this.

On the other hand, the group of GC candidates with redder colours show a detectable spatial concentration towards NGC\,4546. In this sense, it cannot be ruled out that these objects are an extension of the classic red GCs, but follow-up spectroscopy is required to definitely prove whether this is the case. However, it is noteworthy that most of these redder objects have similar behaviour in colour planes involving $g'$, $r'$, $i'$ and $z'$ bands as some GC candidates studied by \citet{powalka16}. These authors, using different colour-colour diagrams in the $ugrizK_s$ filters, found a few tens of GCs in the Virgo cluster core region (some confirmed with radial velocity), which show different behavior to the bulk of their GCs sample when $g'$, $r'$, $i'$ and $z'$ bands are used. Some of these objects are clearly located on the MW stellar sequence in the $(r'-z')$ vs $(g'-r')$ colour-colour diagram. Interesting, the authors also found that the MW GCs share the same locus in this kind of colour plane. While these objects are uniformly spread over the cluster region, it is important to note that some of them are grouped primarily around the lenticular galaxy M86 and the peculiar spiral galaxy NGC\,4438. As mentioned by \citet{powalka16}, it is then plausible that these GCs are born in an environment less dense than the Virgo core, which could be what we observed in NGC\,4546.
However, it is still necessary to consider different options that could explain the very red colours of these objects in NGC\,4546. Some options that would explain this could be due to a reddening effect by dust in the galaxy, or objects belonging to an old GC subpopulation with higher metallicity, and/or objects associated with star clusters called faint fuzzies \citep[FF;][]{larsen00,forbes14}.
When comparing the location of these objects around NGC\,4546 (see bottom panel in Figure \ref{figure_11}) and the dust substructures (see Section \ref{sec:ellipse}), it can be seen that only a few of them are found near these regions. Therefore, this effect would not be the main cause of the presence of group E. Regarding the FF, these objects are generally associated with the disks of S0 galaxies located in low-density environments \citep{liu16}, showing large sizes (effective radii of $7-15$ pc), low luminosity ($M_V>-7$), usually with red colours, and also with high metallicities and old ages. Some of the features shown by the objects belonging to group E in NGC\,4546 (detectable spatial concentration; low brightness $g'_0>23$ mag or $M_V>-7.5$; very red colours) seem to agree with those of the FF. Unfortunately, due to the lack of spectroscopic data of these objects and the impossibility of measuring their radii with our ground-based imaging due to the distance to NGC\,4546, it is not possible to determine their metallicities and/or their sizes in our GMOS images to confirm its nature.

Returning to intermediate GC candidates, the analysis of their GCLF showed that the TO value obtained for this subpopulation is marginally brighter than the values corresponding to blue and red GC subpopulations. This feature could indicate younger objects and/or with different metallicities in comparison to the aforementioned subpopulations. However, since these GC candidates do not constitute a dominant and significantly bright population in the colour-magnitude diagram, as seen in other recent galaxy mergers \citep[e.g., GCs of $\sim2$ Gyr in NGC\,1316;][]{sesto16}, they would not be objects very young nor massive. In contrast, it is necessary to mention that the galaxy does not seem to show evidence of a recent star formation. \citet{kuntschner10} conducted a spectroscopic study of the stellar component of NGC\,4546 using the integral-field spectrograph SAURON. These authors found an age of $11.7^{+1.1}_{-1.0}$ Gyr for the galaxy, using the information obtained by the Lick indices up to $R_\mathrm{eff}$. Also, \citet{mcdermid15} used the same data as these authors but with the full spectral fitting technique, obtaining a mass-weighted age of $13.61\pm0.68$ Gyr. In this sense, the presence of these intermediate GC candidates would suggest that they would not have been formed in a recent star formation event ($<3$ Gyr), given a contribution of young field stars would be expected during that process \citep{forte14}. 

If this intermediate colour GC group is really different from the classic subpopulations, a possible scenario about its origin could involve the young NGC\,4546-UCD1 mentioned above. \citet{norris15} assume that this object is the final result of tidal interaction between NGC\,4546 and a dwarf galaxy of mass $\sim3\times10^{9}$\, M$_\odot$. NGC\,4546-UCD1, with a metallicity $\mathrm{[Z/H]}=0.18$ dex and an extended star formation history from early epochs has previously interacted with the galaxy in the last Gyrs, beginning its current state between $1-3$ Gyr ago. These previous encounters ($>3$ Gyr) may have formed new GCs with a different metallicity compared with the typical blue and red ones, which we observed in our analysis. In this context, if we assume that the intermediate GC subpopulation could have originated as a result of this interaction, using single stellar population photometric models, their colour peak at $(g'-z')_0\sim1.15$ mag would indicate a mean age and metallicity of $\sim$5 Gyr and $\mathrm{[Z/H]}\sim-0.3$ dex, respectively. 

Regarding the origin of lenticular galaxies in low-density environments, \citet{bekki11} have shown that encounters, interactions and unequal mergers of gas-rich disks in group environments could transform late-type galaxies in lenticulars. According to their simulation, it is expected that new stars form during the process of transformation. Therefore, the possible detection of intermediate colour GCs in this galaxy could be the most obvious indication of a strong transformation process that happened some Gyr ago, and highlight the potentiality of GC studies to trace the violent past of their host galaxies. In order to confirm it a spectroscopic study of the GCs subpopulation and a more detailed analysis of the stellar population of field stars looking for younger stellar components in NGC\,4546 is necessary.

%===========================================================================
%===========================================================================

\section*{Acknowledgments}
We thank the anonymous referee for his/her constructive comments.
This work was funded with grants from Consejo Nacional de Investigaciones
Cientificas y Tecnicas de la Republica Argentina, and Universidad Nacional
de La Plata (Argentina). Based on observations obtained at the Gemini Observatory, 
which is operated by the Association of Universities for Research in Astronomy, Inc., 
under a cooperative agreement with the NSF on behalf of the Gemini partnership: the 
National Science Foundation (United States), the National Research Council (Canada), 
CONICYT (Chile), the Australian Research Council (Australia), Minist\'{e}rio da 
Ci\^{e}ncia, Tecnologia e Inova\c{c}\~{a}o (Brazil) and Ministerio de Ciencia, 
Tecnolog\'{i}a e Innovaci\'{o}n Productiva (Argentina). The Gemini program ID are GS-2011A-Q-13 and GS-2014A-Q-30. 
This research has made use of the NED, which is operated by the Jet Propulsion Laboratory, 
Caltech, under contract with the National Aeronautics and Space Administration. 

\bibliographystyle{mnras}
\bibliography{bibliography_4546}

\begin{thebibliography}{}
\makeatletter
\relax
\def\mn@urlcharsother{\let\do\@makeother \do\$\do\&\do\#\do\^\do\_\do\%\do\~}
\def\mn@doi{\begingroup\mn@urlcharsother \@ifnextchar [ {\mn@doi@}
  {\mn@doi@[]}}
\def\mn@doi@[#1]#2{\def\@tempa{#1}\ifx\@tempa\@empty \href
  {http://dx.doi.org/#2} {doi:#2}\else \href {http://dx.doi.org/#2} {#1}\fi
  \endgroup}
\def\mn@eprint#1#2{\mn@eprint@#1:#2::\@nil}
\def\mn@eprint@arXiv#1{\href {http://arxiv.org/abs/#1} {{\tt arXiv:#1}}}
\def\mn@eprint@dblp#1{\href {http://dblp.uni-trier.de/rec/bibtex/#1.xml}
  {dblp:#1}}
\def\mn@eprint@#1:#2:#3:#4\@nil{\def\@tempa {#1}\def\@tempb {#2}\def\@tempc
  {#3}\ifx \@tempc \@empty \let \@tempc \@tempb \let \@tempb \@tempa \fi \ifx
  \@tempb \@empty \def\@tempb {arXiv}\fi \@ifundefined
  {mn@eprint@\@tempb}{\@tempb:\@tempc}{\expandafter \expandafter \csname
  mn@eprint@\@tempb\endcsname \expandafter{\@tempc}}}

\bibitem[\protect\citeauthoryear{{Akaike}}{{Akaike}}{1974}]{akaike74}
{Akaike} H.,  1974, IEEE Transactions on Automatic Control, \href
  {http://adsabs.harvard.edu/abs/1974ITAC...19..716A} {19, 716}

\bibitem[\protect\citeauthoryear{{Baron}}{{Baron}}{2019}]{baron19}
{Baron} D.,  2019, arXiv e-prints, \href
  {https://ui.adsabs.harvard.edu/abs/2019arXiv190407248B} {p. arXiv:1904.07248}

\bibitem[\protect\citeauthoryear{{Beasley}, {Trujillo}, {Leaman}  \&
  {Montes}}{{Beasley} et~al.}{2018}]{beasley18}
{Beasley} M.~A.,  {Trujillo} I.,  {Leaman} R.,   {Montes} M.,  2018, \mn@doi
  [\nat] {10.1038/nature25756}, \href
  {https://ui.adsabs.harvard.edu/abs/2018Natur.555..483B} {555, 483}

\bibitem[\protect\citeauthoryear{{Bekki} \& {Couch}}{{Bekki} \&
  {Couch}}{2011}]{bekki11}
{Bekki} K.,  {Couch} W.~J.,  2011, \mn@doi [\mnras]
  {10.1111/j.1365-2966.2011.18821.x}, \href
  {https://ui.adsabs.harvard.edu/abs/2011MNRAS.415.1783B} {415, 1783}

\bibitem[\protect\citeauthoryear{{Bekki}, {Yahagi}, {Nagashima}  \&
  {Forbes}}{{Bekki} et~al.}{2008}]{bekki08}
{Bekki} K.,  {Yahagi} H.,  {Nagashima} M.,   {Forbes} D.~A.,  2008, \mn@doi
  [\mnras] {10.1111/j.1365-2966.2008.13318.x}, \href
  {http://adsabs.harvard.edu/abs/2008MNRAS.387.1131B} {387, 1131}

\bibitem[\protect\citeauthoryear{{Bertin} \& {Arnouts}}{{Bertin} \&
  {Arnouts}}{1996}]{bertin96}
{Bertin} E.,  {Arnouts} S.,  1996, \mn@doi [\aaps] {10.1051/aas:1996164}, \href
  {http://adsabs.harvard.edu/abs/1996A%26AS..117..393B} {117, 393}

\bibitem[\protect\citeauthoryear{{Bettoni}, {Galletta}  \&
  {Oosterloo}}{{Bettoni} et~al.}{1991}]{bettoni91}
{Bettoni} D.,  {Galletta} G.,   {Oosterloo} T.,  1991, \mn@doi [\mnras]
  {10.1093/mnras/248.3.544}, \href
  {http://adsabs.harvard.edu/abs/1991MNRAS.248..544B} {248, 544}

\bibitem[\protect\citeauthoryear{{Blom}, {Spitler}  \& {Forbes}}{{Blom}
  et~al.}{2012}]{blom12}
{Blom} C.,  {Spitler} L.~R.,   {Forbes} D.~A.,  2012, \mn@doi [\mnras]
  {10.1111/j.1365-2966.2011.19963.x}, \href
  {http://adsabs.harvard.edu/abs/2012MNRAS.420...37B} {420, 37}

\bibitem[\protect\citeauthoryear{{Bonfini}, {Zezas}, {Birkinshaw}, {Worrall},
  {Fabbiano}, {O'Sullivan}, {Trinchieri}  \& {Wolter}}{{Bonfini}
  et~al.}{2012}]{bonfini12}
{Bonfini} P.,  {Zezas} A.,  {Birkinshaw} M.,  {Worrall} D.~M.,  {Fabbiano} G.,
  {O'Sullivan} E.,  {Trinchieri} G.,   {Wolter} A.,  2012, \mn@doi [\mnras]
  {10.1111/j.1365-2966.2012.20514.x}, \href
  {http://adsabs.harvard.edu/abs/2012MNRAS.421.2872B} {421, 2872}

\bibitem[\protect\citeauthoryear{{Bressan}, {Marigo}, {Girardi}, {Salasnich},
  {Dal Cero}, {Rubele}  \& {Nanni}}{{Bressan} et~al.}{2012}]{bressan12}
{Bressan} A.,  {Marigo} P.,  {Girardi} L.,  {Salasnich} B.,  {Dal Cero} C.,
  {Rubele} S.,   {Nanni} A.,  2012, \mn@doi [\mnras]
  {10.1111/j.1365-2966.2012.21948.x}, \href
  {http://ads.astro.puc.cl/abs/2012MNRAS.427..127B} {427, 127}

\bibitem[\protect\citeauthoryear{{Bridges} et~al.,}{{Bridges}
  et~al.}{2006}]{bridges06}
{Bridges} T.,  et~al., 2006, \mn@doi [\mnras]
  {10.1111/j.1365-2966.2006.10997.x}, \href
  {http://adsabs.harvard.edu/abs/2006MNRAS.373..157B} {373, 157}

\bibitem[\protect\citeauthoryear{{Brodie} \& {Strader}}{{Brodie} \&
  {Strader}}{2006}]{brodie06}
{Brodie} J.~P.,  {Strader} J.,  2006, \mn@doi [\araa]
  {10.1146/annurev.astro.44.051905.092441}, \href
  {http://adsabs.harvard.edu/abs/2006ARA%26A..44..193B} {44, 193}

\bibitem[\protect\citeauthoryear{{Cappellari} et~al.,}{{Cappellari}
  et~al.}{2011}]{cappellari11}
{Cappellari} M.,  et~al., 2011, \mn@doi [\mnras]
  {10.1111/j.1365-2966.2010.18174.x}, \href
  {http://adsabs.harvard.edu/abs/2011MNRAS.413..813C} {413, 813}

\bibitem[\protect\citeauthoryear{{Cappellari} et~al.,}{{Cappellari}
  et~al.}{2013}]{cappellari13}
{Cappellari} M.,  et~al., 2013, \mn@doi [\mnras] {10.1093/mnras/stt562}, \href
  {http://adsabs.harvard.edu/abs/2013MNRAS.432.1709C} {432, 1709}

\bibitem[\protect\citeauthoryear{{Caso}, {Richtler}, {Bassino}, {Salinas},
  {Lane}  \& {Romanowsky}}{{Caso} et~al.}{2013}]{caso13}
{Caso} J.~P.,  {Richtler} T.,  {Bassino} L.~P.,  {Salinas} R.,  {Lane} R.~R.,
  {Romanowsky} A.,  2013, \mn@doi [\aap] {10.1051/0004-6361/201321032}, \href
  {http://adsabs.harvard.edu/abs/2013A%26A...555A..56C} {555, A56}

\bibitem[\protect\citeauthoryear{{Caso}, {Bassino}  \& {G{\'o}mez}}{{Caso}
  et~al.}{2015}]{caso15}
{Caso} J.~P.,  {Bassino} L.~P.,   {G{\'o}mez} M.,  2015, \mn@doi [\mnras]
  {10.1093/mnras/stv2015}, \href
  {http://adsabs.harvard.edu/abs/2015MNRAS.453.4421C} {453, 4421}

\bibitem[\protect\citeauthoryear{{Choksi}, {Gnedin}  \& {Li}}{{Choksi}
  et~al.}{2018}]{choksi18}
{Choksi} N.,  {Gnedin} O.~Y.,   {Li} H.,  2018, \mn@doi [\mnras]
  {10.1093/mnras/sty1952}, \href
  {https://ui.adsabs.harvard.edu/abs/2018MNRAS.480.2343C} {480, 2343}

\bibitem[\protect\citeauthoryear{{Colless} et~al.,}{{Colless}
  et~al.}{2003}]{colless03}
{Colless} M.,  et~al., 2003, ArXiv Astrophysics e-prints, \href
  {http://adsabs.harvard.edu/abs/2003astro.ph..6581C} {}

\bibitem[\protect\citeauthoryear{{D'Abrusco}, {Fabbiano}, {Mineo}, {Strader},
  {Fragos}, {Kim}, {Luo}  \& {Zezas}}{{D'Abrusco} et~al.}{2014}]{dabrusco14}
{D'Abrusco} R.,  {Fabbiano} G.,  {Mineo} S.,  {Strader} J.,  {Fragos} T.,
  {Kim} D.-W.,  {Luo} B.,   {Zezas} A.,  2014, \mn@doi [\apj]
  {10.1088/0004-637X/783/1/18}, \href
  {http://adsabs.harvard.edu/abs/2014ApJ...783...18D} {783, 18}

\bibitem[\protect\citeauthoryear{{Eliche-Moral}, {Rodr{\'{\i}}guez-P{\'e}rez},
  {Borlaff}, {Querejeta}  \& {Tapia}}{{Eliche-Moral} et~al.}{2018}]{eliche18}
{Eliche-Moral} M.~C.,  {Rodr{\'{\i}}guez-P{\'e}rez} C.,  {Borlaff} A.,
  {Querejeta} M.,   {Tapia} T.,  2018, \mn@doi [\aap]
  {10.1051/0004-6361/201832911}, \href
  {http://adsabs.harvard.edu/abs/2018A%26A...617A.113E} {617, A113}

\bibitem[\protect\citeauthoryear{{Emsellem} et~al.,}{{Emsellem}
  et~al.}{2004}]{emsellem04}
{Emsellem} E.,  et~al., 2004, \mn@doi [\mnras]
  {10.1111/j.1365-2966.2004.07948.x}, \href
  {http://adsabs.harvard.edu/abs/2004MNRAS.352..721E} {352, 721}

\bibitem[\protect\citeauthoryear{{Erben} et~al.,}{{Erben}
  et~al.}{2005}]{erben05}
{Erben} T.,  et~al., 2005, \mn@doi [Astronomische Nachrichten]
  {10.1002/asna.200510396}, \href
  {http://adsabs.harvard.edu/abs/2005AN....326..432E} {326, 432}

\bibitem[\protect\citeauthoryear{{Escudero}, {Faifer}, {Bassino},
  {Calder{\'o}n}  \& {Caso}}{{Escudero} et~al.}{2015}]{escudero15}
{Escudero} C.~G.,  {Faifer} F.~R.,  {Bassino} L.~P.,  {Calder{\'o}n} J.~P.,
  {Caso} J.~P.,  2015, \mn@doi [\mnras] {10.1093/mnras/stv283}, \href
  {http://adsabs.harvard.edu/abs/2015MNRAS.449..612E} {449, 612}

\bibitem[\protect\citeauthoryear{{Escudero}, {Faifer}, {Smith Castelli},
  {Forte}, {Sesto}, {Gonz{\'a}lez}  \& {Scalia}}{{Escudero}
  et~al.}{2018}]{escudero18}
{Escudero} C.~G.,  {Faifer} F.~R.,  {Smith Castelli} A.~V.,  {Forte} J.~C.,
  {Sesto} L.~A.,  {Gonz{\'a}lez} N.~M.,   {Scalia} M.~C.,  2018, \mn@doi
  [\mnras] {10.1093/mnras/stx3045}, \href
  {http://adsabs.harvard.edu/abs/2018MNRAS.474.4302E} {474, 4302}

\bibitem[\protect\citeauthoryear{{Faifer} et~al.,}{{Faifer}
  et~al.}{2011}]{faifer11}
{Faifer} F.~R.,  et~al., 2011, \mn@doi [\mnras]
  {10.1111/j.1365-2966.2011.19018.x}, \href
  {http://adsabs.harvard.edu/abs/2011MNRAS.416..155F} {416, 155}

\bibitem[\protect\citeauthoryear{{Forbes} \& {Bridges}}{{Forbes} \&
  {Bridges}}{2010}]{forbes10}
{Forbes} D.~A.,  {Bridges} T.,  2010, \mn@doi [\mnras]
  {10.1111/j.1365-2966.2010.16373.x}, \href
  {https://ui.adsabs.harvard.edu/abs/2010MNRAS.404.1203F} {404, 1203}

\bibitem[\protect\citeauthoryear{{Forbes}, {Almeida}, {Spitler}  \&
  {Pota}}{{Forbes} et~al.}{2014}]{forbes14}
{Forbes} D.~A.,  {Almeida} A.,  {Spitler} L.~R.,   {Pota} V.,  2014, \mn@doi
  [\mnras] {10.1093/mnras/stu940}, \href
  {http://adsabs.harvard.edu/abs/2014MNRAS.442.1049F} {442, 1049}

\bibitem[\protect\citeauthoryear{{Forte}, {Vega}, {Faifer}, {Smith Castelli},
  {Escudero}, {Gonz{\'a}lez}  \& {Sesto}}{{Forte} et~al.}{2014}]{forte14}
{Forte} J.~C.,  {Vega} E.~I.,  {Faifer} F.~R.,  {Smith Castelli} A.~V.,
  {Escudero} C.,  {Gonz{\'a}lez} N.~M.,   {Sesto} L.,  2014, \mn@doi [\mnras]
  {10.1093/mnras/stu658}, \href
  {http://adsabs.harvard.edu/abs/2014MNRAS.441.1391F} {441, 1391}

\bibitem[\protect\citeauthoryear{{Forte} et~al.,}{{Forte}
  et~al.}{2019}]{forte19}
{Forte} J.~C.,  et~al., 2019, \mn@doi [\mnras] {10.1093/mnras/sty2746}, \href
  {https://ui.adsabs.harvard.edu/abs/2019MNRAS.482..950F} {482, 950}

\bibitem[\protect\citeauthoryear{{Fukugita}, {Shimasaku}  \&
  {Ichikawa}}{{Fukugita} et~al.}{1995}]{fukugita95}
{Fukugita} M.,  {Shimasaku} K.,   {Ichikawa} T.,  1995, \mn@doi [\pasp]
  {10.1086/133643}, \href {http://adsabs.harvard.edu/abs/1995PASP..107..945F}
  {107, 945}

\bibitem[\protect\citeauthoryear{{Fukugita}, {Ichikawa}, {Gunn}, {Doi},
  {Shimasaku}  \& {Schneider}}{{Fukugita} et~al.}{1996}]{fukugita96}
{Fukugita} M.,  {Ichikawa} T.,  {Gunn} J.~E.,  {Doi} M.,  {Shimasaku} K.,
  {Schneider} D.~P.,  1996, \mn@doi [\aj] {10.1086/117915}, \href
  {http://adsabs.harvard.edu/abs/1996AJ....111.1748F} {111, 1748}

\bibitem[\protect\citeauthoryear{{Galletta}}{{Galletta}}{1987}]{galletta87}
{Galletta} G.,  1987, \mn@doi [\apj] {10.1086/165389}, \href
  {http://adsabs.harvard.edu/abs/1987ApJ...318..531G} {318, 531}

\bibitem[\protect\citeauthoryear{{Gao}, {Ho}, {Barth}  \& {Li}}{{Gao}
  et~al.}{2018}]{gao18}
{Gao} H.,  {Ho} L.~C.,  {Barth} A.~J.,   {Li} Z.-Y.,  2018, \mn@doi [\apj]
  {10.3847/1538-4357/aacdac}, \href
  {http://adsabs.harvard.edu/abs/2018ApJ...862..100G} {862, 100}

\bibitem[\protect\citeauthoryear{{Georgiev}, {Puzia}, {Goudfrooij}  \&
  {Hilker}}{{Georgiev} et~al.}{2010}]{georgiev10}
{Georgiev} I.~Y.,  {Puzia} T.~H.,  {Goudfrooij} P.,   {Hilker} M.,  2010,
  \mn@doi [\mnras] {10.1111/j.1365-2966.2010.16802.x}, \href
  {https://ui.adsabs.harvard.edu/abs/2010MNRAS.406.1967G} {406, 1967}

\bibitem[\protect\citeauthoryear{{Girardi}, {Groenewegen}, {Hatziminaoglou}  \&
  {da Costa}}{{Girardi} et~al.}{2005}]{girardi05}
{Girardi} L.,  {Groenewegen} M.~A.~T.,  {Hatziminaoglou} E.,   {da Costa} L.,
  2005, \mn@doi [\aap] {10.1051/0004-6361:20042352}, \href
  {http://ads.astro.puc.cl/abs/2005A%26A...436..895G} {436, 895}

\bibitem[\protect\citeauthoryear{{Harris}}{{Harris}}{2009}]{harris09a}
{Harris} W.~E.,  2009, \mn@doi [\apj] {10.1088/0004-637X/699/1/254}, \href
  {http://adsabs.harvard.edu/abs/2009ApJ...699..254H} {699, 254}

\bibitem[\protect\citeauthoryear{{Harris} \& {van den Bergh}}{{Harris} \& {van
  den Bergh}}{1981}]{harris81}
{Harris} W.~E.,  {van den Bergh} S.,  1981, \mn@doi [\aj] {10.1086/113047},
  \href {http://adsabs.harvard.edu/abs/1981AJ.....86.1627H} {86, 1627}

\bibitem[\protect\citeauthoryear{{Harris}, {Harris}  \& {Alessi}}{{Harris}
  et~al.}{2013}]{harris13}
{Harris} W.~E.,  {Harris} G.~L.~H.,   {Alessi} M.,  2013, \mn@doi [\apj]
  {10.1088/0004-637X/772/2/82}, \href
  {http://adsabs.harvard.edu/abs/2013ApJ...772...82H} {772, 82}

\bibitem[\protect\citeauthoryear{{Harris}, {Blakeslee}  \& {Harris}}{{Harris}
  et~al.}{2017}]{harris17}
{Harris} W.~E.,  {Blakeslee} J.~P.,   {Harris} G. L.~H.,  2017, \mn@doi [\apj]
  {10.3847/1538-4357/836/1/67}, \href
  {https://ui.adsabs.harvard.edu/abs/2017ApJ...836...67H} {836, 67}

\bibitem[\protect\citeauthoryear{{Ho}, {Li}, {Barth}, {Seigar}  \& {Peng}}{{Ho}
  et~al.}{2011}]{ho11}
{Ho} L.~C.,  {Li} Z.-Y.,  {Barth} A.~J.,  {Seigar} M.~S.,   {Peng} C.~Y.,
  2011, \mn@doi [\apjs] {10.1088/0067-0049/197/2/21}, \href
  {http://adsabs.harvard.edu/abs/2011ApJS..197...21H} {197, 21}

\bibitem[\protect\citeauthoryear{{Hook}, {J{\o}rgensen}, {Allington-Smith},
  {Davies}, {Metcalfe}, {Murowinski}  \& {Crampton}}{{Hook}
  et~al.}{2004}]{hook04}
{Hook} I.~M.,  {J{\o}rgensen} I.,  {Allington-Smith} J.~R.,  {Davies} R.~L.,
  {Metcalfe} N.,  {Murowinski} R.~G.,   {Crampton} D.,  2004, \mn@doi [\pasp]
  {10.1086/383624}, \href {http://adsabs.harvard.edu/abs/2004PASP..116..425H}
  {116, 425}

\bibitem[\protect\citeauthoryear{{Huang}, {Ho}, {Peng}, {Li}  \&
  {Barth}}{{Huang} et~al.}{2013}]{huang13}
{Huang} S.,  {Ho} L.~C.,  {Peng} C.~Y.,  {Li} Z.-Y.,   {Barth} A.~J.,  2013,
  \mn@doi [\apj] {10.1088/0004-637X/766/1/47}, \href
  {http://adsabs.harvard.edu/abs/2013ApJ...766...47H} {766, 47}

\bibitem[\protect\citeauthoryear{{Jord{\'a}n} et~al.,}{{Jord{\'a}n}
  et~al.}{2007}]{jordan07}
{Jord{\'a}n} A.,  et~al., 2007, \mn@doi [\apj] {10.1086/516840}, \href
  {http://adsabs.harvard.edu/abs/2007ApJS..171..101J} {171, 101}

\bibitem[\protect\citeauthoryear{{J{\o}rgensen}}{{J{\o}rgensen}}{2009}]{jorgensen09}
{J{\o}rgensen} I.,  2009, \mn@doi [\pasa] {10.1071/AS08008}, \href
  {http://adsabs.harvard.edu/abs/2009PASA...26...17J} {26, 17}

\bibitem[\protect\citeauthoryear{{Kissler-Patig}, {Forbes}  \&
  {Minniti}}{{Kissler-Patig} et~al.}{1998}]{kissler98}
{Kissler-Patig} M.,  {Forbes} D.~A.,   {Minniti} D.,  1998, \mn@doi [\mnras]
  {10.1046/j.1365-8711.1998.01725.x}, \href
  {http://adsabs.harvard.edu/abs/1998MNRAS.298.1123K} {298, 1123}

\bibitem[\protect\citeauthoryear{{Kruijssen}}{{Kruijssen}}{2014}]{kruijssen14}
{Kruijssen} J.~M.~D.,  2014, \mn@doi [Classical and Quantum Gravity]
  {10.1088/0264-9381/31/24/244006}, \href
  {http://ads.astro.puc.cl/abs/2014CQGra..31x4006K} {31, 244006}

\bibitem[\protect\citeauthoryear{{Kruijssen}}{{Kruijssen}}{2015}]{kruijssen15}
{Kruijssen} J.~M.~D.,  2015, \mn@doi [\mnras] {10.1093/mnras/stv2026}, \href
  {http://adsabs.harvard.edu/abs/2015MNRAS.454.1658K} {454, 1658}

\bibitem[\protect\citeauthoryear{{Kuntschner}, {Ziegler}, {Sharples}, {Worthey}
   \& {Fricke}}{{Kuntschner} et~al.}{2002}]{kuntschner02}
{Kuntschner} H.,  {Ziegler} B.~L.,  {Sharples} R.~M.,  {Worthey} G.,   {Fricke}
  K.~J.,  2002, \mn@doi [\aap] {10.1051/0004-6361:20021325}, \href
  {http://adsabs.harvard.edu/abs/2002A%26A...395..761K} {395, 761}

\bibitem[\protect\citeauthoryear{{Kuntschner} et~al.,}{{Kuntschner}
  et~al.}{2010}]{kuntschner10}
{Kuntschner} H.,  et~al., 2010, \mn@doi [\mnras]
  {10.1111/j.1365-2966.2010.17161.x}, \href
  {http://adsabs.harvard.edu/abs/2010MNRAS.408...97K} {408, 97}

\bibitem[\protect\citeauthoryear{{Lane}, {Salinas}  \& {Richtler}}{{Lane}
  et~al.}{2013}]{lane13}
{Lane} R.~R.,  {Salinas} R.,   {Richtler} T.,  2013, \mn@doi [\aap]
  {10.1051/0004-6361/201220231}, \href
  {http://adsabs.harvard.edu/abs/2013A%26A...549A.148L} {549, A148}

\bibitem[\protect\citeauthoryear{{Larsen} \& {Brodie}}{{Larsen} \&
  {Brodie}}{2000}]{larsen00}
{Larsen} S.~S.,  {Brodie} J.~P.,  2000, \mn@doi [\aj] {10.1086/316847}, \href
  {http://adsabs.harvard.edu/abs/2000AJ....120.2938L} {120, 2938}

\bibitem[\protect\citeauthoryear{{Lee}, {Park}, {Hwang}, {Arimoto}, {Tamura}
  \& {Onodera}}{{Lee} et~al.}{2010}]{lee10}
{Lee} M.~G.,  {Park} H.~S.,  {Hwang} H.~S.,  {Arimoto} N.,  {Tamura} N.,
  {Onodera} M.,  2010, \mn@doi [\apj] {10.1088/0004-637X/709/2/1083}, \href
  {http://ads.astro.puc.cl/abs/2010ApJ...709.1083L} {709, 1083}

\bibitem[\protect\citeauthoryear{{Li}, {Ho}, {Barth}  \& {Peng}}{{Li}
  et~al.}{2011}]{yuli11}
{Li} Z.-Y.,  {Ho} L.~C.,  {Barth} A.~J.,   {Peng} C.~Y.,  2011, \mn@doi [\apjs]
  {10.1088/0067-0049/197/2/22}, \href
  {http://adsabs.harvard.edu/abs/2011ApJS..197...22L} {197, 22}

\bibitem[\protect\citeauthoryear{{Liu}, {Peng}, {Lim}, {Jord{\'a}n},
  {Blakeslee}, {C{\^o}t{\'e}}, {Ferrarese}  \& {Pattarakijwanich}}{{Liu}
  et~al.}{2016}]{liu16}
{Liu} Y.,  {Peng} E.~W.,  {Lim} S.,  {Jord{\'a}n} A.,  {Blakeslee} J.,
  {C{\^o}t{\'e}} P.,  {Ferrarese} L.,   {Pattarakijwanich} P.,  2016, \mn@doi
  [\apj] {10.3847/0004-637X/830/2/99}, \href
  {https://ui.adsabs.harvard.edu/abs/2016ApJ...830...99L} {830, 99}

\bibitem[\protect\citeauthoryear{{McDermid} et~al.,}{{McDermid}
  et~al.}{2015}]{mcdermid15}
{McDermid} R.~M.,  et~al., 2015, \mn@doi [\mnras] {10.1093/mnras/stv105}, \href
  {http://adsabs.harvard.edu/abs/2015MNRAS.448.3484M} {448, 3484}

\bibitem[\protect\citeauthoryear{{McLaughlin}, {Harris}  \&
  {Hanes}}{{McLaughlin} et~al.}{1994}]{mclaughlin94}
{McLaughlin} D.~E.,  {Harris} W.~E.,   {Hanes} D.~A.,  1994, \mn@doi [\apj]
  {10.1086/173744}, \href {http://adsabs.harvard.edu/abs/1994ApJ...422..486M}
  {422, 486}

\bibitem[\protect\citeauthoryear{{Metcalfe}, {Shanks}, {Campos}, {McCracken}
  \& {Fong}}{{Metcalfe} et~al.}{2001}]{metcalfe01}
{Metcalfe} N.,  {Shanks} T.,  {Campos} A.,  {McCracken} H.~J.,   {Fong} R.,
  2001, \mn@doi [\mnras] {10.1046/j.1365-8711.2001.04168.x}, \href
  {http://adsabs.harvard.edu/abs/2001MNRAS.323..795M} {323, 795}

\bibitem[\protect\citeauthoryear{{Mieske}, {Hilker}, {Infante}  \&
  {Jord{\'a}n}}{{Mieske} et~al.}{2006}]{mieske06b}
{Mieske} S.,  {Hilker} M.,  {Infante} L.,   {Jord{\'a}n} A.,  2006, \mn@doi
  [\aj] {10.1086/500583}, \href
  {http://adsabs.harvard.edu/abs/2006AJ....131.2442M} {131, 2442}

\bibitem[\protect\citeauthoryear{{Norris} \& {Kannappan}}{{Norris} \&
  {Kannappan}}{2011}]{norris11}
{Norris} M.~A.,  {Kannappan} S.~J.,  2011, \mn@doi [\mnras]
  {10.1111/j.1365-2966.2011.18440.x}, \href
  {http://adsabs.harvard.edu/abs/2011MNRAS.414..739N} {414, 739}

\bibitem[\protect\citeauthoryear{{Norris}, {Escudero}, {Faifer}, {Kannappan},
  {Forte}  \& {van den Bosch}}{{Norris} et~al.}{2015}]{norris15}
{Norris} M.~A.,  {Escudero} C.~G.,  {Faifer} F.~R.,  {Kannappan} S.~J.,
  {Forte} J.~C.,   {van den Bosch} R.~C.~E.,  2015, \mn@doi [\mnras]
  {10.1093/mnras/stv1221}, \href
  {http://adsabs.harvard.edu/abs/2015MNRAS.451.3615N} {451, 3615}

\bibitem[\protect\citeauthoryear{{Peng}, {Ho}, {Impey}  \& {Rix}}{{Peng}
  et~al.}{2002}]{peng02}
{Peng} C.~Y.,  {Ho} L.~C.,  {Impey} C.~D.,   {Rix} H.-W.,  2002, \mn@doi [\aj]
  {10.1086/340952}, \href {http://adsabs.harvard.edu/abs/2002AJ....124..266P}
  {124, 266}

\bibitem[\protect\citeauthoryear{{Peng} et~al.,}{{Peng} et~al.}{2006}]{peng06}
{Peng} E.~W.,  et~al., 2006, \mn@doi [\apj] {10.1086/498210}, \href
  {http://adsabs.harvard.edu/abs/2006ApJ...639...95P} {639, 95}

\bibitem[\protect\citeauthoryear{{Peng}, {Ho}, {Impey}  \& {Rix}}{{Peng}
  et~al.}{2010}]{peng10}
{Peng} C.~Y.,  {Ho} L.~C.,  {Impey} C.~D.,   {Rix} H.-W.,  2010, \mn@doi [\aj]
  {10.1088/0004-6256/139/6/2097}, \href
  {http://adsabs.harvard.edu/abs/2010AJ....139.2097P} {139, 2097}

\bibitem[\protect\citeauthoryear{{Pierce} et~al.,}{{Pierce}
  et~al.}{2006a}]{pierce06a}
{Pierce} M.,  et~al., 2006a, \mn@doi [\mnras]
  {10.1111/j.1365-2966.2005.09810.x}, \href
  {http://adsabs.harvard.edu/abs/2006MNRAS.366.1253P} {366, 1253}

\bibitem[\protect\citeauthoryear{{Pierce} et~al.,}{{Pierce}
  et~al.}{2006b}]{pierce06b}
{Pierce} M.,  et~al., 2006b, \mn@doi [\mnras]
  {10.1111/j.1365-2966.2006.10097.x}, \href
  {http://adsabs.harvard.edu/abs/2006MNRAS.368..325P} {368, 325}

\bibitem[\protect\citeauthoryear{{Powalka} et~al.,}{{Powalka}
  et~al.}{2016}]{powalka16}
{Powalka} M.,  et~al., 2016, \mn@doi [\apj] {10.3847/2041-8205/829/1/L5}, \href
  {https://ui.adsabs.harvard.edu/abs/2016ApJ...829L...5P} {829, L5}

\bibitem[\protect\citeauthoryear{{Puzia} et~al.,}{{Puzia}
  et~al.}{2004}]{puzia04}
{Puzia} T.~H.,  et~al., 2004, \mn@doi [\aap] {10.1051/0004-6361:20031448},
  \href {http://adsabs.harvard.edu/abs/2004A%26A...415..123P} {415, 123}

\bibitem[\protect\citeauthoryear{{Puzia}, {Kissler-Patig}, {Thomas},
  {Maraston}, {Saglia}, {Bender}, {Goudfrooij}  \& {Hempel}}{{Puzia}
  et~al.}{2005}]{puzia05}
{Puzia} T.~H.,  {Kissler-Patig} M.,  {Thomas} D.,  {Maraston} C.,  {Saglia}
  R.~P.,  {Bender} R.,  {Goudfrooij} P.,   {Hempel} M.,  2005, \mn@doi [\aap]
  {10.1051/0004-6361:20047012}, \href
  {http://adsabs.harvard.edu/abs/2005A%26A...439..997P} {439, 997}

\bibitem[\protect\citeauthoryear{{Puzia}, {Paolillo}, {Goudfrooij},
  {Maccarone}, {Fabbiano}  \& {Angelini}}{{Puzia} et~al.}{2014}]{puzia14}
{Puzia} T.~H.,  {Paolillo} M.,  {Goudfrooij} P.,  {Maccarone} T.~J.,
  {Fabbiano} G.,   {Angelini} L.,  2014, \mn@doi [\apj]
  {10.1088/0004-637X/786/2/78}, \href
  {http://adsabs.harvard.edu/abs/2014ApJ...786...78P} {786, 78}

\bibitem[\protect\citeauthoryear{{Ricci}, {Steiner}  \& {Menezes}}{{Ricci}
  et~al.}{2015}]{ricci15}
{Ricci} T.~V.,  {Steiner} J.~E.,   {Menezes} R.~B.,  2015, \mn@doi [\mnras]
  {10.1093/mnras/stv1156}, \href
  {http://adsabs.harvard.edu/abs/2015MNRAS.451.3728R} {451, 3728}

\bibitem[\protect\citeauthoryear{{Salinas}, {Alabi}, {Richtler}  \&
  {Lane}}{{Salinas} et~al.}{2015}]{salinas15}
{Salinas} R.,  {Alabi} A.,  {Richtler} T.,   {Lane} R.~R.,  2015, \mn@doi
  [\aap] {10.1051/0004-6361/201425574}, \href
  {http://adsabs.harvard.edu/abs/2015A%26A...577A..59S} {577, A59}

\bibitem[\protect\citeauthoryear{{Schirmer}}{{Schirmer}}{2013}]{schirmer13}
{Schirmer} M.,  2013, \mn@doi [\apjs] {10.1088/0067-0049/209/2/21}, \href
  {http://adsabs.harvard.edu/abs/2013ApJS..209...21S} {209, 21}

\bibitem[\protect\citeauthoryear{{Schlafly} \& {Finkbeiner}}{{Schlafly} \&
  {Finkbeiner}}{2011}]{schlafly11}
{Schlafly} E.~F.,  {Finkbeiner} D.~P.,  2011, \mn@doi [\apj]
  {10.1088/0004-637X/737/2/103}, \href
  {http://adsabs.harvard.edu/abs/2011ApJ...737..103S} {737, 103}

\bibitem[\protect\citeauthoryear{{S\'ersic}}{{S\'ersic}}{1968}]{sersic68}
{S\'ersic} J.~L.,  1968, {Atlas de galaxias australes}

\bibitem[\protect\citeauthoryear{{Sesto}, {Faifer}  \& {Forte}}{{Sesto}
  et~al.}{2016}]{sesto16}
{Sesto} L.~A.,  {Faifer} F.~R.,   {Forte} J.~C.,  2016, \mn@doi [\mnras]
  {10.1093/mnras/stw1627}, \href
  {http://adsabs.harvard.edu/abs/2016MNRAS.461.4260S} {461, 4260}

\bibitem[\protect\citeauthoryear{{Sesto}, {Faifer}, {Smith Castelli}, {Forte}
  \& {Escudero}}{{Sesto} et~al.}{2018}]{sesto18}
{Sesto} L.~A.,  {Faifer} F.~R.,  {Smith Castelli} A.~V.,  {Forte} J.~C.,
  {Escudero} C.~G.,  2018, \mn@doi [\mnras] {10.1093/mnras/sty1416}, \href
  {http://adsabs.harvard.edu/abs/2018MNRAS.479..478S} {479, 478}

\bibitem[\protect\citeauthoryear{{Sheth} et~al.,}{{Sheth}
  et~al.}{2010}]{sheth10}
{Sheth} K.,  et~al., 2010, \mn@doi [\pasp] {10.1086/657638}, \href
  {http://ads.astro.puc.cl/abs/2010PASP..122.1397S} {122, 1397}

\bibitem[\protect\citeauthoryear{{Stetson}}{{Stetson}}{1987}]{stetson87}
{Stetson} P.~B.,  1987, \mn@doi [\pasp] {10.1086/131977}, \href
  {http://adsabs.harvard.edu/abs/1987PASP...99..191S} {99, 191}

\bibitem[\protect\citeauthoryear{{Strader}, {Brodie}, {Cenarro}, {Beasley}  \&
  {Forbes}}{{Strader} et~al.}{2005}]{strader05}
{Strader} J.,  {Brodie} J.~P.,  {Cenarro} A.~J.,  {Beasley} M.~A.,   {Forbes}
  D.~A.,  2005, \mn@doi [\aj] {10.1086/432717}, \href
  {http://adsabs.harvard.edu/abs/2005AJ....130.1315S} {130, 1315}

\bibitem[\protect\citeauthoryear{{Strader}, {Beasley}  \& {Brodie}}{{Strader}
  et~al.}{2007}]{strader07}
{Strader} J.,  {Beasley} M.~A.,   {Brodie} J.~P.,  2007, \mn@doi [\aj]
  {10.1086/512770}, \href {http://adsabs.harvard.edu/abs/2007AJ....133.2015S}
  {133, 2015}

\bibitem[\protect\citeauthoryear{{Tonini}}{{Tonini}}{2013}]{tonini13}
{Tonini} C.,  2013, \mn@doi [\apj] {10.1088/0004-637X/762/1/39}, \href
  {http://adsabs.harvard.edu/abs/2013ApJ...762...39T} {762, 39}

\bibitem[\protect\citeauthoryear{{Tully} et~al.,}{{Tully}
  et~al.}{2013}]{tully13}
{Tully} R.~B.,  et~al., 2013, \mn@doi [\aj] {10.1088/0004-6256/146/4/86}, \href
  {http://adsabs.harvard.edu/abs/2013AJ....146...86T} {146, 86}

\bibitem[\protect\citeauthoryear{{Usher} et~al.,}{{Usher}
  et~al.}{2012}]{usher12}
{Usher} C.,  et~al., 2012, \mn@doi [\mnras] {10.1111/j.1365-2966.2012.21801.x},
  \href {http://ads.astro.puc.cl/abs/2012MNRAS.426.1475U} {426, 1475}

\bibitem[\protect\citeauthoryear{{Usher}, {Forbes}, {Spitler}, {Brodie},
  {Romanowsky}, {Strader}  \& {Woodley}}{{Usher} et~al.}{2013}]{usher13}
{Usher} C.,  {Forbes} D.~A.,  {Spitler} L.~R.,  {Brodie} J.~P.,  {Romanowsky}
  A.~J.,  {Strader} J.,   {Woodley} K.~A.,  2013, \mn@doi [\mnras]
  {10.1093/mnras/stt1637}, \href
  {https://ui.adsabs.harvard.edu/abs/2013MNRAS.436.1172U} {436, 1172}

\bibitem[\protect\citeauthoryear{{Woodley}, {Harris}, {Puzia}, {G{\'o}mez},
  {Harris}  \& {Geisler}}{{Woodley} et~al.}{2010}]{woodley10}
{Woodley} K.~A.,  {Harris} W.~E.,  {Puzia} T.~H.,  {G{\'o}mez} M.,  {Harris}
  G.~L.~H.,   {Geisler} D.,  2010, \mn@doi [\apj]
  {10.1088/0004-637X/708/2/1335}, \href
  {http://adsabs.harvard.edu/abs/2010ApJ...708.1335W} {708, 1335}

\bibitem[\protect\citeauthoryear{{Zaritsky} et~al.,}{{Zaritsky}
  et~al.}{2015}]{zaritsky15}
{Zaritsky} D.,  et~al., 2015, \mn@doi [\apj] {10.1088/0004-637X/799/2/159},
  \href {http://ads.astro.puc.cl/abs/2015ApJ...799..159Z} {799, 159}

\bibitem[\protect\citeauthoryear{{Zepf} \& {Ashman}}{{Zepf} \&
  {Ashman}}{1993}]{zepf93}
{Zepf} S.~E.,  {Ashman} K.~M.,  1993, \mn@doi [\mnras]
  {10.1093/mnras/264.3.611}, \href
  {https://ui.adsabs.harvard.edu/abs/1993MNRAS.264..611Z} {264, 611}

\bibitem[\protect\citeauthoryear{{de Souza} et~al.,}{{de Souza}
  et~al.}{2017}]{desouza17}
{de Souza} R.~S.,  et~al., 2017, \mn@doi [\mnras] {10.1093/mnras/stx2156},
  \href {http://adsabs.harvard.edu/abs/2017MNRAS.472.2808D} {472, 2808}

\makeatother
\end{thebibliography}

\end{document}